\documentclass[aps, prb,twocolumn,superscriptaddress]{revtex4-1}
\usepackage{graphicx}
\usepackage{amsmath}
\usepackage{epstopdf}

\usepackage[abs]{overpic}
\usepackage{subfigure}
\usepackage[dvipsnames,svgnames]{xcolor}

\newcommand{\Wigner}{Wigner Research Center for Physics and Optics, Hungarian Academy of Sciences, P.O. Box 49, H-1525 Budapest, Hungary}
\newcommand{\BPUni}{Department of Atomic Physics, Budapest University of Technology and Economics, Budafoki\'{u}t 8., H-1111 Budapest, Hungary}

\begin{document}

\title{Characterization of oxygen defects in diamond by means of density functional theory calculations
}

\author{Gerg\H{o} Thiering}
\affiliation{\Wigner}
\affiliation{\BPUni}
\author{Adam Gali}
\affiliation{\Wigner} 
\affiliation{\BPUni}

\date{\today}

\begin{abstract} 
Point defects in diamond are of high interest as candidates for realizing solid state quantum bits, bioimaging agents, or ultrasensitive electric or magnetic field sensors. Various artificial diamond synthesis methods should introduce oxygen contamination in diamond, however, the incorporation of oxygen into diamond crystal and the nature of oxygen-related point defects are largely unknown. Oxygen may be potentially interesting as a source of quantum bits or it may interact with other point defects which are well established solid state qubits. Here we employ plane-wave supercell calculations within density functional theory, in order to characterize the electronic and magneto-optical properties of various oxygen-related defects. Beside the trivial single interstitial and substitutional oxygen defects we also consider their complexes with vacancies and hydrogen atoms. We find that oxygen defects are mostly electrically active and introduce highly correlated orbitals that pose a challenge for density functional theory modeling. Nevertheless, we are able to identify the fingerprints of substitutional oxygen defect, the oxygen-vacancy and oxygen-vacancy-hydrogen complexes in the electron paramagnetic resonance spectrum. We demonstrate that first principles calculations can predict the motional averaging of the electron paramagnetic resonance spectrum of defects that are subject to Jahn-Teller distortion. We show that the high-spin neutral oxygen-vacancy defect exhibits very fast non-radiative decay from its optical excited state that might hinder to apply it as a qubit.
\end{abstract}

\maketitle

\section{Introduction}

Perfect diamond is a transparent and diamagnetic material. Impurities in diamond may form paramagnetic color centers when they introduce defects states in the fundamental band gap. These defects are of high interest and candidates to realize solid state single photon sources and quantum bits. The impurities often enter diamond in the growth process of diamond. In this study we focus our attention to oxygen impurity in diamond. Previous investigations showed \cite{Han_CVD_improve_1997, Chang_CVD_improve_1992,Wolden_CVD_improve_1996, Tang2004} that addition of oxygen precursors in the diamond chemical vapor deposition (CVD) synthesis process improves the growth speed and the quality of diamond. The formation of diamond from carbon and hydrogen precursors in the presence of oxygen occurs when the number of oxygen and carbon atoms is about the same according to the C, H, O phase diagram \cite{bachmann1991towards}. Oxygen may enter diamond in this process. As oxygen is a light element and the natural abundance of its spinless $^{16}\mathrm{O}$ isotope is close to 100\%,  it is very challenging to observe the fingerprints of oxygen impurity in diamond by vibrational or electron spin resonance (ESR) spectroscopy techniques. Nevertheless, two ESR centers, labeled as WAR5 and OVH, have been tentatively associated with oxygen-related defects in such CVD samples \cite{cann2009magnetic_WAR5, hartland2014study}. WAR5 ESR center exhibits $S=1$ spin state with $C_{3v}$ symmetry and anisotropic $g_{\perp}$=2.0026(2)  $g_{||}$=2.0029(2) $g$-tensor where $||$ and $\perp$ refer to direction parallel and perpendicular to the $\langle 111 \rangle$ symmetry axis, respectively. The measured zero-field-splitting $D$ constant\cite{NV_D1, NV_D2, cann2009magnetic_WAR5} and the hyperfine couplings between the electron spin and proximate $^{13}$C nuclear spins (natural abundance of 1.1\%)\cite{NV_1977hyper, NVhyper_PhysRevB.77.155206, cann2009magnetic_WAR5} was very similar to those of the negatively charged nitrogen-vacancy color center, NV($-$) \cite{Preez_phd_1965, Davies1976}, in diamond. As the electronic structure of NV($-$) and OV($0$) is isovalent, WAR5 was proposed to originate from OV($0$) \cite{cann2009magnetic_WAR5}. We note that the formation of WAR5 center was accompanied by a 543.2-nm photoluminescence (PL) center \cite{cann2009magnetic_WAR5}. However, photoexcitation of the diamond sample containing the WAR5 ESR and 543.2-nm PL centers did not show any spinpolarization of the WAR5 ESR center that does not share the well-known spinpolarization properties of NV($-$) center \cite{Doherty2013}. It has been not yet understood whether these ESR and PL centers have no common origin or they have common origin but the spinpolarization effect in OV center is negligible for other reasons. The other oxygen-related OVH ESR center has $S=1/2$ spin state with a characteristic $^1$H hyperfine signal of $A_{||}$=$\pm$13.6(1)~MHz, $A_{\perp}$=$\mp$9.0(1)~MHz and $g$-tensor of $g_{||}$=2.0034(1), $g_{\perp}$=2.0029(1) \cite{hartland2014study}. This ESR center was very analogous with the negatively charged NVH($-$) ESR center \cite{Glover2003}, thus it was proposed that the OVH ESR center should originate from the neutral OVH($0$) defect because these two defects are isovalent. In addition, this OVH ESR center was annealed out at 1800$^{\circ}$C where NVH($-$) defect anneals too \cite{Glover2003} that further strengthens the association of OVH ESR center with the OVH($0$) defect. We emphasize that the association of the WAR5 and OVH ESR centers with OV($0$) and OVH($0$) defects, respectively, seems plausible but no direct evidence of the presence of oxygen impurity has been demonstrated in these two ESR centers. We further note that a series of new cathodoluminescence (CL) lines were found in the region of 455.0~nm and 484.1~nm in CVD diamonds where 0.1\% O$_2$ contamination was used in the CVD gas mixture, and they were tentatively associated with oxygen defects \cite{Ruan1.108685}. The difficulties of identification of oxygen-related defects may be represented by the case of the OK1 \cite{Newton1989} and N3 \cite{Wyk1992} ESR centers, where these centers were first tentatively assigned to oxygen-related defects in type Ib natural diamonds (see Ref.~\onlinecite{Etmimi2010} and references therein), however, later it has been shown by further ESR and other studies that they are rather related to Ti defects \cite{Nadolinny2009, Nadolinny2012}.

Oxygen may be introduced by other synthesis techniques. It was reported that oxygen-acetylene flame in ambient atmosphere deposited on various substrates results in diamond\cite{Hanssen_torch_1988} where the 543.2-nm PL center was reported \cite{543nm_1990,543nm_1991} that was also found in CVD diamonds with H:C:O chemistry \cite{cann2009magnetic_WAR5}. When CVD and high-pressure high-temperature (HPHT) diamonds were produced in the presence of CrO$_3$ then CL centers at 330~nm and 267~nm were speculatively assigned to oxygen defects \cite{Mori1.107368}. Defects tentatively associated with oxygen were also reported in diamonds synthesized from HPHT treatment of $\mathrm{Na_2CO_3\!-\!CO_2\!-\!C}$ carbonate medium at pressure and temperature of 6.3~GPa and 1673-1773~K. In these diamonds the OX1, OX2, OX3 $S=1/2$ ESR centers emerged \cite{Komarovskikh2013} with $g$ factors of $g_1=g_2=2.0032(1)$, $g_3=2.0038(1)$;  $g_{1}=2.0098(1)$, $g_{2}=1.9991(1)$, $g_{3}=2.0113(1)$ ;  $g_{1}=2.0064(1)$, $g_{2}=2.0008(1)$,  $g_{3}=2.0052(1)$ respectively. While the OX2, OX3 did not have any observable hyperfine features, the OX1 center seems to possess hyperfine satellites at $A_{1}$=914~MHz and $A_{2}$=729~MHz $A_{3}$=589~MHz that were associated with $^{13}$C nuclear spin. Diamonds synthesized in $\mathrm{Na_2CO_3\!-\!CO_2\!-\!H_2O\!-\!C}$ filament with the same temperature and pressure conditions a new $S=1/2$ ESR center was found that was labeled as VOH \cite{Komarovskikh2014}. This ESR center was fitted with $g_{||}$=2.0023(1), $g_{\perp}$=2.0030(1), and $A_{||}$35(1)~MHz,  $A_{\perp}$=32(1)~MHz spin-Hamiltonian parameters where the hyperfine coupling originates from an $I=1/2$ nuclear spin with $\approx$100\% abundance that was associated with a single $\mathrm{^1H}$ \cite{Komarovskikh2014}. This ESR center clearly differs from the one measured in CVD samples which is labeled as OVH \cite{hartland2014study}. In order not to confuse the readers, we use only OVH label in the context exclusively for the ESR center in CVD samples. The participation of oxygen impurity in all of these four point defects in the special HPHT diamonds was proposed because of the relatively large anisotropy of the $g$-factor of the defects. The anisotropy of the $g$-factor was assumed \cite{Komarovskikh2014} to originate from the large spin-orbit coupling $\lambda$ of the oxygen atom ($\lambda$=151~cm$^{-1}$) characteristic to this atom type.

The oxygen may be definitely introduced into diamond by oxygen ion implantation. Oxygen-ion implantation was argued to lead to \textit{n}-type conduction with activation energy 0.32 eV~in type IIa diamond samples,\cite{Prins2000} however, no independent experimental study has yet confirmed this claim. PL centers associated with oxygen impurity were reported \cite{Gippius1993640} in natural type IIa and Ia diamond samples at 584.8, 598.3, 845.0 and 836.0~nm after $^{16}\mathrm{O}$ ion-implantation. These PL centers appeared after annealing at 1500~$^\circ$C unlike other PL centers formed by the same bombardments \cite{Gippius1993640}, nevertheless, the assignment is still speculative. An unambiguous signal from oxygen-related defect was reported in diamond samples implanted by $^{17}\mathrm{O}$ ions \cite{KUL12}. The $^{17}\mathrm{O}$ isotopes have $I=5/2$ nuclear spin, thus its interaction with electron spins is basically observable in the ESR spectrum. Indeed, KUL12 ESR center with $S=1/2$ spin and isotropic $g$=2.0023(1) factor was found in $^{17}\mathrm{O}$ implanted, nitrogen-free CVD diamond films where the $A_{\perp}$=207(3)~MHz and  $A_{||}$=238(3)~MHz hyperfine satellite lines were associated with the presence of $^{17}\mathrm{O}$ isotope. This ESR center is stable even after annealing at 1400~$^{\circ}$C. The microscopic origin of the KUL12 ESR center has not yet been identified. The properties of the hyperfine and $g$ tensors imply a high symmetry defect.  The few theoretical studies on oxygen defects in diamond focused on the electronic structure and ionization energies \cite{Gali2001, Lowther2003, Goss2004, Goss2005}, thus the origin of this center is still unraveled. 

In summary, the knowledge about the incorporation and the role of oxygen impurities in diamond is scarce despite the fact that oxygen is presumably introduced in various synthesis processes and can be engineered by oxygen ion implantation. As a first step in understanding of oxygen impurities in diamond, we carry out a systematic study on single oxygen related defects and their complexes with vacancy and hydrogen impurities by means of plane wave supercell density functional theory calculations. Beside electronic structure calculations, we determine the magneto-optical properties of the considered defects, and compare the calculated and experimental data for identification of oxygen-related spectroscopy centers.
The paper is organized as follows. In the following Sec.~\ref{sec:Methodology} we briefly describe the applied methodology. We discuss the results of atomistic simulations on interstitial (Sec.~\ref{sec:Oi}), substitutional (Sec.~\ref{sec:Os}) oxygen defects, as well as their complexes with vacancy (Sec.~\ref{sec:OV}) and hydrogen impurities (Sec.~\ref{sec:OH}). In the neutral substitional oxygen defect and the complexes of oxygen and vacancy highly correlated orbitals appear that is discussed in conjunction with the applied density functional theory in Sections~\ref{sec:Os} and \ref{sec:OV}, respectively. We study the presence of motional averaging in the ESR spectrum of defects that exhibit Jahn-Teller distortion in Sec.~\ref{sec:DJT}. We analyze the optical properties of the neutral oxygen-vacancy complex in a separate section (Sec.~\ref{sec:543.2}) where we discuss the non-radiative decay process from the optically excited state of the defect, and the connection of the 543.2~nm PL center and the WAR5 ESR center. Finally, we conclude our paper in Sec.~\ref{sec:Conclusion}.

\section{Methodology \label{sec:Methodology}}
We characterize point defects consisting of single oxygen, vacancy and hydrogen atoms in diamond within \emph{ab initio} spin-polarized density functional theory (DFT) as implemented in the \textsc{vasp} code\cite{Kresse1994, Kresse1996}.  We determine the total energy of the considered defects within Born-Oppenheimer approximation: the ions are treated as classical particles and the potential energy surface (PES) is calculated as a function of the position of the ions in the system. The lowest energy in PES belongs to the ground state geometry configuration.  We embed the defects in a cubic 512-atom supercell which is large enough to sample the Brillouin zone only at the $\Gamma$ point. The electron wave functions are expanded in a plane wave basis set. We employ the standard projector augmented-wave (PAW) method \cite{Blchl1994} that allows a relatively low-energy cutoff for the expansion of plane waves and yet accurately provides the charge and spin density in the interstitial region and regions around the ions.

The applied DFT functionals and plane-wave cutoffs are optimized for the considered physical quantities and computational capacity. We applied the Perdew-Burke-Ernzerhof (PBE) generalized gradient approximated variant of DFT \cite{perdew1996generalized} to calculate the ground-state charge and spin densities of the system, as well as quasilocal vibration (phonon) frequencies of the systems. The choice of PBE for calculating the vibrations is motivated by two reasons: (i)  PBE very accurately reproduces the experimental diamond lattice constant, phonon spectrum, and the dependence of these properties of pressure or temperature; \cite{ivanova2013ab} (ii) as we allow all the atoms to vibrate it would be computationally prohibitive to apply non-local functionals to obtain the results. In the \textsc{vasp} implementation within density functional perturbation theory\cite{baroni2001phonons} the calculated Raman-mode at 1326~cm$^{-1}$ in our 512-atom supercell is in very good agreement with the experimental data of 1332~cm$^{-1}$.\cite{nemanich1979first} In the defective calculations, we allowed all the atoms to relax until the forces fall below the threshold of 10$^{-4}$~eV/\AA . In this case we use the PBE optimized lattice constant of diamond (3.565~\AA). The plane-wave energy cutoff is set to 370~eV for the expansion of plane waves and 740~eV for the expansion of the charge density. The atoms are allowed to relax until the forces went below 10$^{-2}$ eV/\AA. We apply this functional to calculate the $D$-tensor of the system \cite{Ivady2014} and the barrier energy of reorientation of some defects where the application of non-local density functional is computationally too demanding.

The charge transition levels as well as the intradefect level optical transition energies are determined by HSE06 \cite{HSE03, HSE06} screened hybrid functional because it provides quantitatively good agreement for defects in diamond \cite{Gali2009, Deak2010, Deak2014, gali2013ab}. In this case the optimized lattice constant of diamond is 3.545~\AA. We employ the same threshold for the forces in the geometry optimization procedure like that in PBE calculations. We apply the $\Delta$SCF method to calculate the excited states of the system that allows us to calculate the PES in the excited state, thus the zero-phonon-line (ZPL) energies can be determined and compared to the experimental data \cite{Gali2009, gali2013ab}. The applied plane-wave cutoff is set to 370~eV except in the calculations of hyperfine couplings between the electron and nuclei spins where the cutoff is raised up to 600~eV. We calculate the hyperfine couplings including the core spin polarization in the Fermi-contact interaction within the frozen valence approximation \cite{yazyev2005core} as implemented in the \textsc{vasp} code \cite{szasz2013hyperfine}. We note that the hard PAW-potentials of oxygen and hydrogen are applied in the calculation of hyperfine tensors on $^{17}$O and $^{1}$H, respectively.

We determine the formation energy [$E^q_\text{f}\left(E_\text{F}\right)$] as a function of Fermi-level $E_\text{F}$ in the fundamental band gap, in charge state $q$ of the defect \cite{Zhang1991}, 
\begin{equation}
\begin{split}E^q_\text{f}\left(E_\text{F}\right)=
E^q_\text{tot}-\sum_{\mathrm{C,O,H}} n_{i}\mu_i-q (E_\text{F} - E_\text{V}) + \Delta E^q_\text{corr} \text{,}
\end{split}
\label{eq:formation}
\end{equation}
where $\mu_\text{i}$ is the chemical potential of the corresponding atom. The $\mu_\text{C}$ can be calculated from the perfect diamond lattice (-10.56~eV). The value of $\mu_\text{O}$ is referenced to the CO molecule in vacuum (-7.71~eV) in the carbon rich limit. Finally the $\mu_\text{H}$ is deduced from a 22-layer slab model of (001)-($2\times1$) reconstructed and hydrogen terminated diamond surface (-3.71~eV) \cite{Deak2014}. In the context of the rest of paper, the origo of the Fermi-level $E_F$ is aligned to the calculated valence band edge $E_\text{V}$ that we consequently apply in the plots. The remaining $\Delta E^q_\text{corr}$ term is a correction in the total energy due of the defective charged supercell. Here, we apply a relatively simple formula, 2/3 of the monopole term of the Makov-Payne correction \cite{Makov1995, Lany2008} after average potential correction, which practically yields equivalent results with the Freysoldt correction \cite{Freysoldt2009, Trinh2013} and is able to properly reproduce the experimental ionization energies of deep defects in group-IV crystals \cite{Deak2010} including the defects in diamond \cite{gali2013ab, Deak2014} within the accuracy of about 0.1~eV.

The adiabatic charge transition levels for a single defect $E(q|q+1)$ can be derived from Eq.~\ref{eq:formation} and are calculated as
\begin{equation}
E(q|q+1)=E_\text{tot}^q-E_\text{tot}^{q+1}+ \Delta E^q_\text{corr}-\Delta E^{q+1}_\text{corr} \text{,}
\label{eq:transition}
\end{equation}
which yields the position of the Fermi level where the formation energy of the defect for the two charge states $q$ and $q+1$ are equal.

We calculate the binding energies with the following formula,
\begin{equation}
E_\text{bind} =     E_\text{f}^{q_{A}} [A] 
							   +E_\text{f}^{q_B} [B] 
							   -E_\text{f}^{q_C} [C] + \sum _{i={A,B,C}} q_i E_\text{F} \text{,}
\label{eq:binding}
\end{equation}
where the reactant is the defect $C$ created by complex formation of $A$ and $B$ isolated defects. In this definition, positive binding energy means favorable reaction and a larger value of $E_\text{bind}$ indicates a larger probability of reaction where this depends on the actual position of $E_\text{F}$ which sets the stable charge states ($q_i$) of the defects in the reaction.

\section{Results}

\begin{figure}[ht]
\includegraphics[width=1\columnwidth]{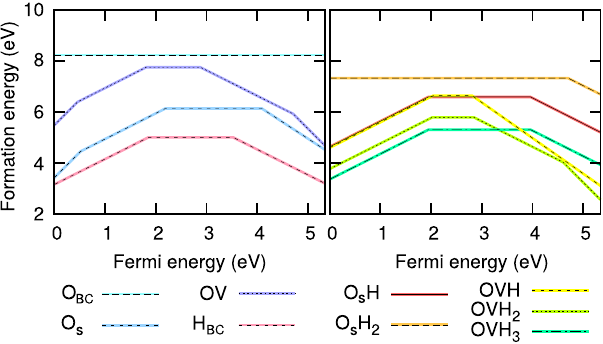}
\caption{(color online). 
Formation energies of the considered oxygen defects and the interstitial hydrogen (H$_\text{BC}$) defect. The chemical potential of oxygen and hydrogen are taken from CO molecule and a hydrogen terminated diamond surface in vacuum, respectively. See text for more details.
}
\label{fig:formation}
\end{figure}

We study the basic oxygen defects, interstitial (O$_\text{BC}$) and substitutional (O$_\text{S}$) oxygen, the complex of oxygen-vacancy (OV) in diamond. In addition, we consider the complex formation of  O$_\text{S}$ with one and two hydrogen atoms (O$_\text{S}$H and O$_\text{S}$H$_2$, respectively) as well as OV with one, two and three hydrogen atoms (OVH, OVH$_2$, and OVH$_3$, respectively). The calculated formation energies are summarized in Fig.~\ref{fig:formation}.

The interstitial oxygen has the highest formation energy and substitutional oxygen is much favorable. The oxygen-vacancy complex is less favorable than O$_\text{S}$, however, the complex formation from O$_\text{S}$ and vacancy is preferred by 5.23~eV (Fermi level is at the acceptor level of the vacancy, E$_\text{V}$+1.9~eV)\cite{Deak2014}. Interstitial hydrogen (H$_\text{BC}$ bond centered configuration in its mobile positively charged and neutral state) may form complexes with  O$_\text{S}$ or OV defects. The calculated binding energy of O$_\text{S}$ and H$_\text{BC}$ is 4.37~eV whereas it is 4.27~eV for O$_\text{S}$H and H$_\text{BC}$ to form O$_\text{S}$H$_2$ when the Fermi level is set to E$_\text{V}$+2.0~eV. Complex formation of OV, OVH, and OVH$_2$ with H$_\text{BC}$ is favorable by 6.13~eV, 5.88~eV, and 5.47~eV, respectively. We note that the formation energies of OV and OVH are similar. This implies that when OH-like radicals occur in the growth process of diamond then OVH may form as a unit during diamond growth.  

In the next Sections we will describe the electronic structure and magneto-optical properties of the considered defects in detail.

\subsection{Interstitial oxygen defect \label{sec:Oi}}

\begin{figure}[ht]
\includegraphics[width=1\columnwidth]{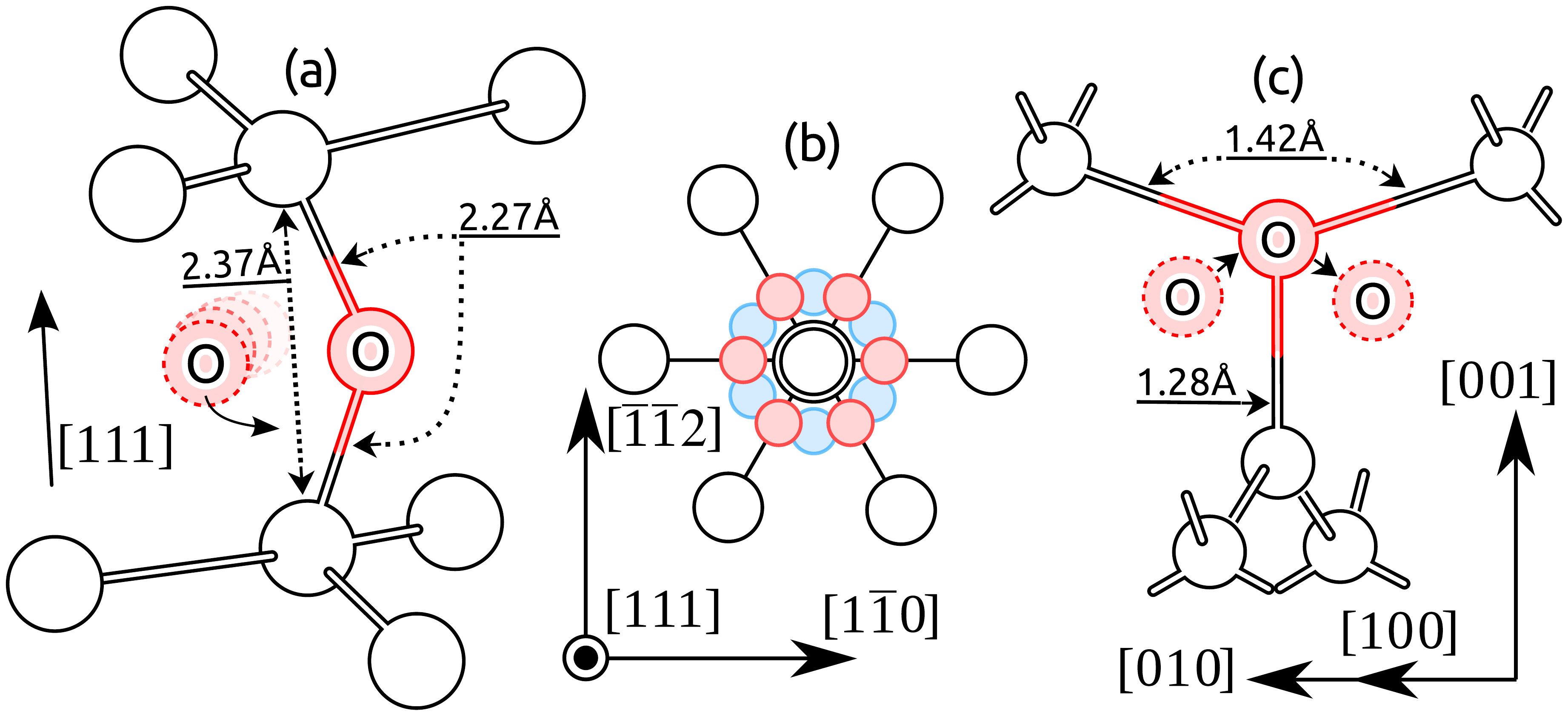}
\caption{(color online). 
 Geometry sketches the interstitial oxygen defects. (a) and (b) describes the bond centered configuration $\mathrm{O_{BC}}$, while (c) depicts the split interstitial defect that is the saddle point of the migration of $\mathrm{O_{BC}}$ interstitial oxygen. The barrier energy for diffusion is 3.13~eV. The blue and red circles represent equivalent distortions in (b).
}
\label{fig:interstitial}
\end{figure}

Initially, we placed an interstitial oxygen atom in the bond centered position,  labeled as $\mathrm{O_{BC}}$. The C-O-C bond is heavily distorted from the [111] symmetry axis, and the oxygen forms a so-called puckered bond centered configuration. As a consequence, the original $D_{3d}$ symmetry is reduced to $C_s$ or $C_2$ symmetry during the geometric relaxation, see red and blue circles in Fig.~\ref{fig:interstitial}(b), respectively. The high $D_{3d}$ symmetry position yields an energy of 0.62~eV higher than that for the puckered bond center configuration. 
No defect state appears in the fundamental band gap of diamond, thus $\mathrm{O_{BC}}$ is electrically and optically inactive with a spin singlet ground state. Our results indicate a motional averaged structure for $\mathrm{O_{BC}}$ defect as depicted in Fig.~\ref{fig:interstitial}(a): the O atom orbits about the symmetry axis. The barrier between the six equivalent distortions [Fig.~\ref{fig:interstitial}(b)] from $D_{3d}$ symmetry is below 8~meV, thus the sombrero shape of PES is axially symmetric within our accuracy. We conclude that the $\mathrm{O_{BC}}$ has a motionally averaged $D_{3d}$ symmetry.

$\mathrm{O_{BC}}$ has high formation energy but may be created by oxygen-ion implantation of diamond that can further migrate in the lattice. The path of diffusion  can be described by the jump between two neighbor $\mathrm{O_{BC}}$ configurations as depicted in Fig.~\ref{fig:interstitial}(c). We determine the PES of the diffusion of $\mathrm{O_{BC}}$ by PBE calculations because of the relatively high cost of hybrid calculations. We find that the total energy only monotonously increases towards the split interstitial configuration resulting in barrier energy of 2.93~eV. By assuming similar PES within HSE06 functional, one can calculate the barrier energy of diffusion by a single calculation constrained at the split interstitial configuration. The calculated saddle point by HSE06 is 3.13~eV. This large barrier energy implies that interstitial oxygen is not mobile in diamond. It is more likely that neutral vacancies \cite{Deak2014} may diffuse to interstitial oxygen and they recombine to $\mathrm{O_{S}}$ defect as the energy gain of this reaction is over 9~eV. 

We note that interstitial oxygen was tentatively proposed as a candidate for OX2 ESR center in oxygenated HPHT diamond \cite{Komarovskikh2013}. Our results exclude the single interstitial oxygen as a good candidate for OX2 center as it has very high formation energy and is not a paramagnetic defect. 

\subsection{Substitutional oxygen defect \label{sec:Os}}

\begin{figure}[ht]
\includegraphics[width=0.95\columnwidth]{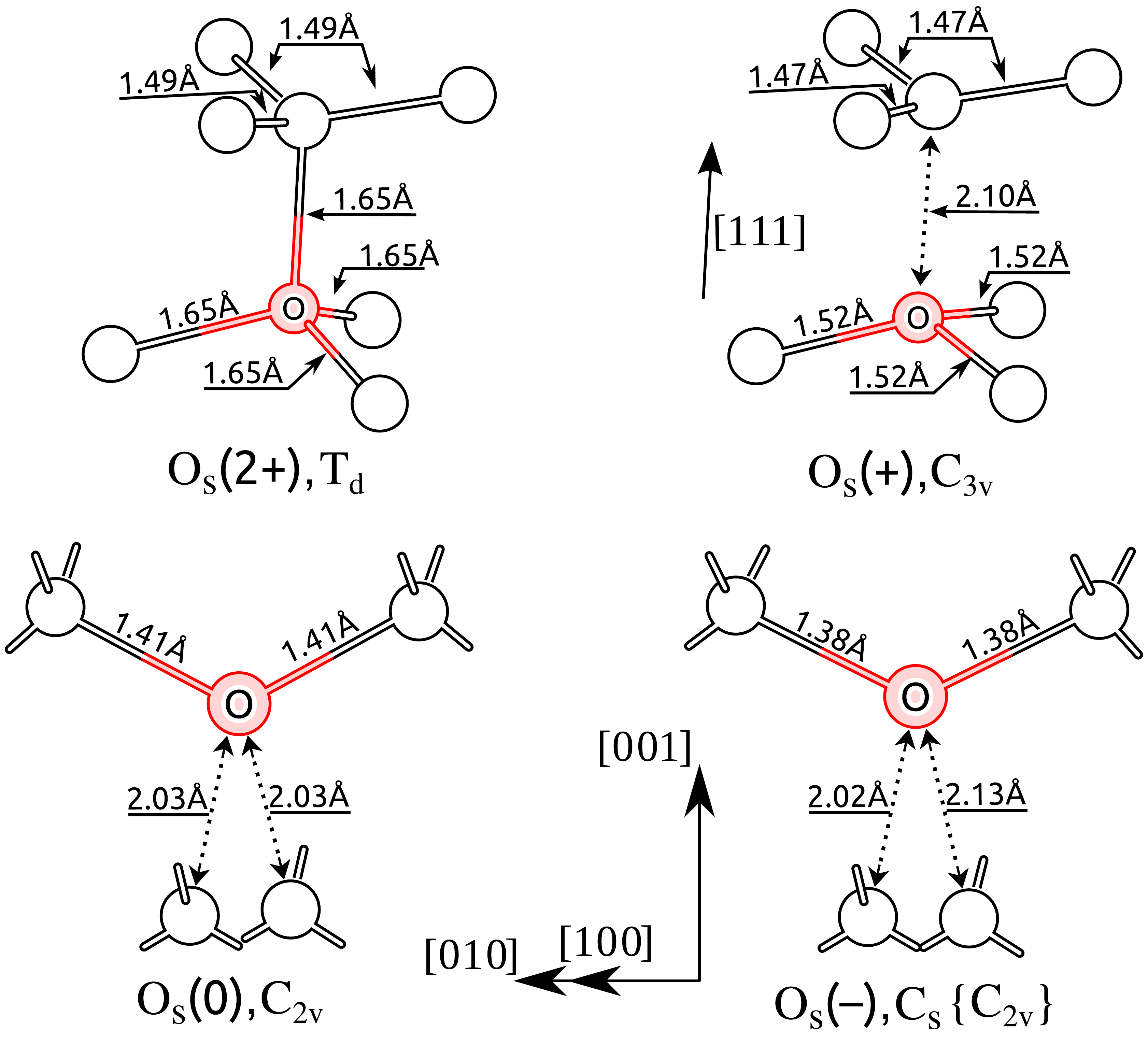}
\caption{(color online). 
Geometry of the $\mathrm{O_\text{S}}$ defect in various charge states. The ($2+$) state is isovalent with carbon resulting in $T_d$ symmetry. By adding electrons to this system those electrons are localized in the dotted elongated bonds in the ($+$) and ($0$) states. Additional electrons do not significantly distort the geometry further.
}
\label{fig:sub}
\end{figure}

\begin{figure}[ht]
\includegraphics[height=160px]{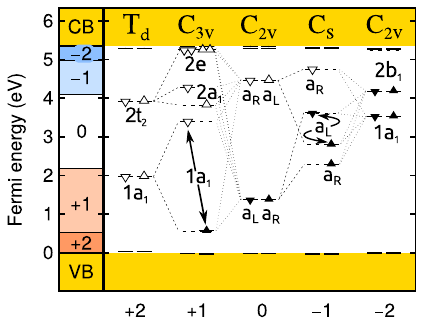}
\caption{(color online). 
Kohn-Sham (KS) levels of the $\mathrm{O_\text{S}}$ defect in various charge states. In the first column the charge transition levels are shown. The symmetry labels of KS states are depicted. The labels $\mathrm{a_R}$ and $\mathrm{a_L}$ are explained in the text and visualized in Fig.~\ref{fig:OsExcitation}.
Yellow area of CB(VB) corresponds to the conduction(valence) band.
}
\label{fig:sub_KS}
\end{figure}

Previous (semi)local DFT calculations already showed \cite{Gali2001, Goss2004} that $\mathrm{O_\text{S}}$ introduces defect levels in the band gap. According to our HSE06 calculations, $\mathrm{O_\text{S}}$  can appear at various charged states depending on the position of the Fermi level. The geometry of the ground state significantly depends on the charge state of the defect as shown in Fig.~\ref{fig:sub}. The driving force in the variance of the geometries in their corresponding charge states is the combination of pseudo Jahn-Teller effect and the substantial correlation of orbitals. The simplest case is the doubly positive charged defect, $\mathrm{O_\text{S}}$($2+$), which is isovalent to the substituted carbon atom. Indeed, that defect possessed the highest $T_d$ point group symmetry with on-center substitution. However, oxygen is much more electronegative than carbon that introduces polarized covalent C-O bonds that are significantly longer (by about 0.11~\AA ) than C-C bonds (1.54~\AA ) in perfect diamond lattice. The occupied defect levels fall in the valence band but the anti-bonding empty $a_1$ and $t_2$ orbitals' levels emerge in the band gap (labeled as $\mathrm{1a_1}$ and $\mathrm{2t_2}$ in Fig.~\ref{fig:sub_KS}).

By adding one electron to this configuration would occupy an $a_1$ antibonding state in $\mathrm{O_\text{S}}$($+$) defect which is principally Jahn-Teller stable. However, distortion from $T_d$ to $C_{3v}$ symmetry results in the splitting of $t_2$ state to $a_1$ and $e$ (labeled as $\mathrm{2a_1}$ and $\mathrm{2e}$ in Fig.~\ref{fig:sub_KS}). That $2a_1$ state can interact with the lower $1a_1$ state with lowering the total energy of the system and drives the system to $C_{3v}$ symmetry. By filling this antibonding orbital with a single electron will break one C-O bond and shortens the remaining three C-O bonds (see Fig.~\ref{fig:sub}). This state has $S=1/2$ electron spin, i.e., it is paramagnetic that may manifest in the electron spin resonance spectrum with the corresponding hyperfine signals. We exclude the feasibility of motional averaging of this defect based on our DFT calculations that will be discussed in a separate section (Sec.~\ref{sec:DJT}). 
The calculated hyperfine tensors of the $\mathrm{^{17}O}$ nuclear $I=5/2$ spin agrees well with the observed hyperfine signal of KUL12 ESR center in $^{17}$O ion implanted diamond samples \cite{KUL12} (see Table~\ref{tab:hyperOs}). As explained above, the stability of $\mathrm{O_\text{S}}$($+$) over $\mathrm{O_{BC}}$ implies that $\mathrm{O_\text{S}}$ indeed form in oxygen implanted diamond. We further note that KUL12 ESR center was detected together with KUL1 ESR center that was identified as the SiV($0$) defect \cite{gali2013ab}. The calculated stability of the neutral SiV, SiV($0$), is in the region of $E_\text{F}=0.27-1.41$~eV that has a common stability window with that of $\mathrm{O_\text{S}}$($+$), i.e., $E_\text{F}=0.32-2.19$~eV (see Fig.~\ref{fig:sub_KS}). All these facts indicate that KUL12 ESR center can be identified as the positively charged substitutional oxygen defect in diamond. 
\begin{table}[htbp]
\caption{Calculated HSE06 hyperfine constants of $\mathrm{O_\text{S}}$($+$) defects. The experimental data is the proposed KUL12 ESR center taken from Ref.~\onlinecite{KUL12}. Only the oxygen hyperfine constants are available for KUL12 ESR center. The direction of the hyperfine constants are given in angles in the parenthesis ($\vartheta$, $\varphi$). $\vartheta$ is the angle from [001], and $\varphi$ is the angle from [100] of the projection into the (001) plane. If no angle is given then the hyperfine interaction is axially symmetric and the high symmetry axis points towards [111]. The direction of $A$ constants are depicted in Fig.~\ref{fig:Os_hyper} with their corresponding labels where the $zz$, $yy$ and $xx$ components are shown by blue, green and red arrows, respectively. The number of symmetrically equivalent $^{13}$C isotopes are indicated in the first column preceding their label, and depicted in Fig.~\ref{fig:hyper_generic}.}
\begin{ruledtabular}
\begin{tabular}{clll}        
     &$A_{zz}$ (MHz) & $A_{yy}$ (MHz) & $A_{xx}$ (MHz) \\
  \hline
KUL12 & & & \\  
 $\mathrm{^{17}O}$ & $\pm$238  & $\pm$207 & $\pm$207 \\  
$\mathrm{O_\text{S}}$($+$) & & &\\       
$\mathrm{^{17}O}$          & -223  & -189 & -189  \\  
   1 (x)  &   314 &  70    & 70   \\    
   3 (a)  &   -29(90,225)&  -27(29,315)   & -24(61,135)  \\    
   3 (c)  &   48(60,45)&  34(150,45)   & 33(90,315)  \\    
   6 (d)   &   13.0(58,294)&  9.0(48,170)   & 8.4(58,6)  \\   
   3 (l)   &   13.5(62,225)&  9.0(90,315)   & 9.0(29,45)  \\   
   other   &  $<4$ &  $<4$    & $<4$   \\ 
\end{tabular} 
\end{ruledtabular}     
\label{tab:hyperOs}
\end{table}
We calculated the hyperfine interaction for proximate $^{13}$C nuclear spins too. A large hyperfine coupling occurs between the carbon dangling bond and the nuclear spin, although, its intensity is only about 0.5\% with respect to the main ESR lines, thus it might be challenging to detect. However, other $^{13}$C detectable hyperfine couplings with 3 or 6 equivalent $^{13}$C nuclear spins with larger intensities can help strengthen the identification of this defect. 

\begin{figure}[ht]
\includegraphics[height=160px]{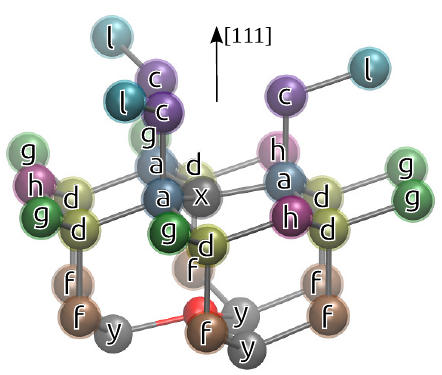}
\caption{(color online). 
Positions of the carbon atoms near the oxygen defect are shown in a perspective view that possess considerable hyperfine interaction. The red-colored non-labeled atom depicts the oxygen atom. Identical colors of balls represent symmetrically equivalent carbon atoms in $C_{3v}$ symmetry. The labels are used in Tables~\ref{tab:hyperOs}, \ref{tab:OV_hyper} and \ref{tab:OVH_hyper}, and Figures \ref{fig:Os_hyper}, \ref{fig:OV_hyper} and \ref{fig:OVH_hyper}. The removal of carbon (x) yields results in OV defect. OVH defect may be created by removing carbon (x) and bonding hydrogen to carbon (a).
}
\label{fig:hyper_generic}
\end{figure}

\begin{figure}[ht]
\includegraphics[height=160px]{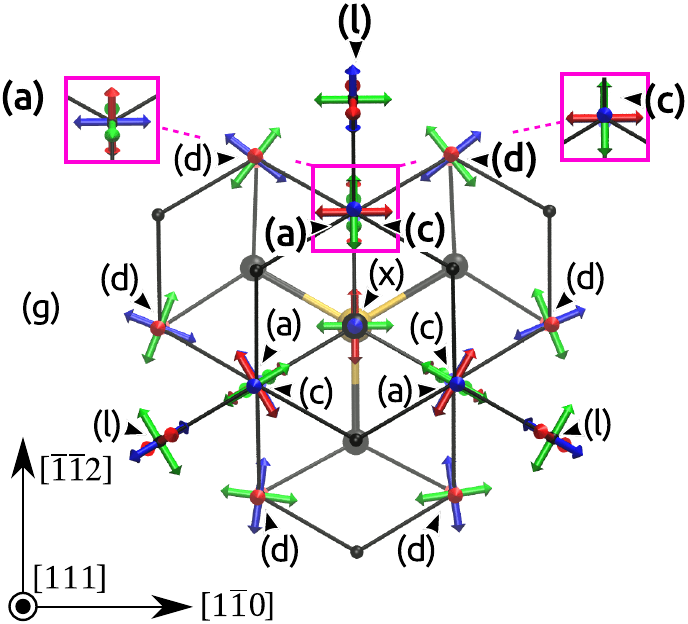}
\caption{(color online). 
The direction of hyperfine eigenvectors of the $\mathrm{O_\text{S}}$($+$) defect are depicted where $zz$, $yy$ and $xx$ components are shown by blue, green and red arrows, respectively. The direction of hyperfine constants labeled by bold letters are in Table~\ref{tab:hyperOs}. The yellow ball represents the O atom. For the sake of clarity, the (a) and (c) hyperfine tensors are shown separately in the two outlying purple boxes. The labeling of carbon atoms agrees with that in Figure \ref{fig:hyper_generic}.
}
\label{fig:Os_hyper}
\end{figure}

\begin{figure*}[ht]
\includegraphics[height=140px]{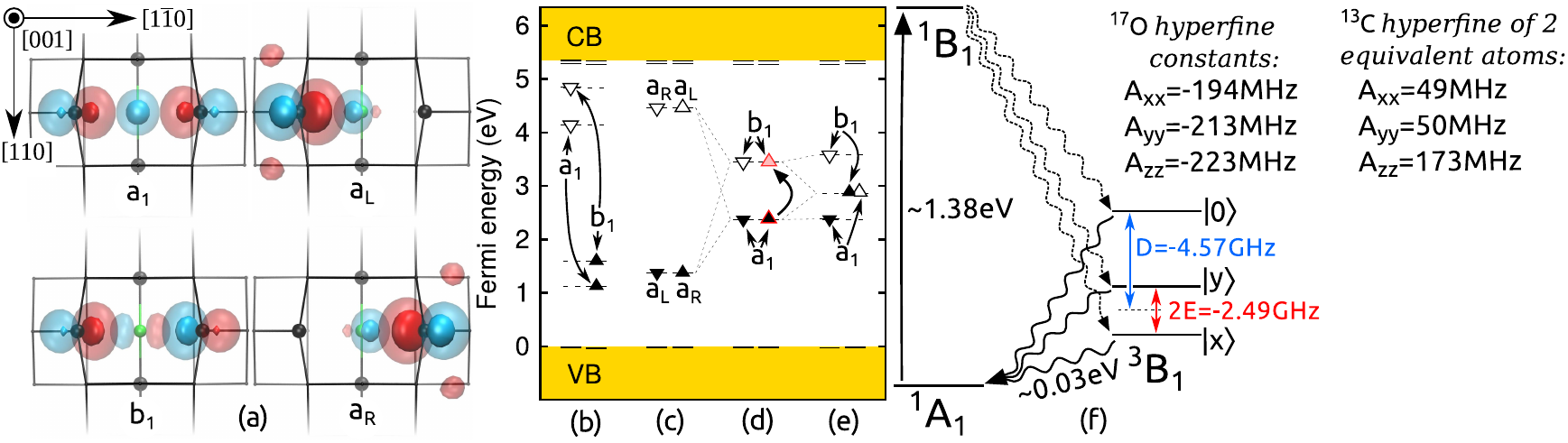}
\caption{(color online). 
Kohn-Sham (KS) states of and the excitation process of $\mathrm{O_\text{S}}$($0$). (a) shows the contour plot of the pseudo-wave functions with isovalues at $\pm 0.006$ (opaque) and $\pm 0.003$ (transparent). The KS energies of the triplet state is shown in (b), whereas (c) corresponds to the symmetry broken singlet KS solution which is lower in total energy by 0.04~eV than that of the high spin case. We show the $C_{2v}$ symmetric singlet KS solution in (d) and the excitation of this system in (e) by promoting one electron from the $a_1$ to $b_1$ level. Yellow area of CB(VB) corresponds to the conduction(valence) band. (f) depicts the many-electron states. The straight arrows represent the optical excitation whereas the wavy dotted(bold) arrows the very weak(efficient) intersystem crossings mediated by spin-orbit and electron-phonon couplings. The calculated zero-field splitting due to electron spin-spin interaction as well as the hyperfine coupling to $^{17}$O nuclear spin are also shown in the triplet state. The hyperfine constants of the two equivalent $^{13}$C atoms are sohwn. All of the remaining $^{13}$C atoms possess hyperfine coupling parameters less than 30 MHz. The spin substates of the triplet is indicated. For the sake of clarity, the energy scale is arbitrary. 
}
\label{fig:OsExcitation}
\end{figure*}

The DFT description of the orbitals of neutral $\mathrm{O_\text{S}}$ is very challenging because it leads to a highly correlated ground state. In the previous (semi)local DFT studies the highly correlated ground state was not considered for  $\mathrm{O_\text{S}}$($0$) \cite{Gali2001, Goss2004}. We showed for SiV-related defects in diamond \cite{Thiering2015} that spinpolarized hybrid density functional theory may be employed to approximate the total energy of highly correlated systems within a 0.2~eV estimated accuracy. We start the discussion of the neutral 
$\mathrm{O_\text{S}}$ with the electronic structure of $\mathrm{O_\text{S}}$($2+$) in $T_d$ symmetry. If two electrons are promoted to the antibonding $1a_1$ and $2t_2$ orbitals then the symmetry will reduce to $C_{2v}$ symmetry because of two broken C-O bonds (see Fig.~\ref{fig:sub}) and two short C-O bonds (1.41~\AA). In other words, oxygen will form two covalent bonds. This configuration often occurs in molecules or defects in solids consisting of an oxygen atom. In this configuration the $2t_2$ state splits to $2b_1$, $2b_2$, and $2a_1$ where the $2b_1$ level lies close to the $1a_1$ level whereas the other levels are resonant with the conduction band edge. This reconstruction may be considered as a pseudo Jahn-Teller effect. According to the \emph{aufbau} principle, the two electrons may fill the $1a_1$ level, $1a_1^2$, resulting in an $^1A_1$ singlet state. However, one should realize that $2b_1^2$ electron configuration has also $^1A_1$ symmetry that can strongly interact with the former state, and they rather form a multideterminant state. The total energy of this multideterminant state can be approximated by symmetry-lowered spinpolarized Kohn-Sham states \cite{Thiering2015} where an antiferromagnetic-like state can be formed. In that state the spin-up electron is localized only in the "right" dangling bond [$\mathrm{a_R}$ state in Fig.~\ref{fig:OsExcitation}(a)] whereas the spin-down electron is localized only in the "left" dangling bond [$\mathrm{a_L}$ state in Fig.~\ref{fig:OsExcitation}(a)]. These Kohn-Sham wave functions show only $C_s$ symmetry that is lower than the $C_{2v}$ symmetry of the structure. This electron configuration has the lowest energy by HSE06 functional. The calculated barrier energy of reorientation between $C_{2v}$ configurations is over 1~eV, thus motional averaged $T_d$ symmetry does not likely occur. 

Since the positively charged substitutional oxygen defect can be definitely generated our theory indicates that the neutral substitutional defect can be found in nitrogen-contaminated diamond. The ground state is singlet, nevertheless, a metastable triplet state may exist and it may be employed as a qubit similar to the ST1 center in diamond \cite{Lee2013}. Therefore, we study the shelving and excited states of $\mathrm{O_\text{S}}$($0$) in detail. We follow a similar methodology as was explained for SiV$_2$:H($-$) defect in diamond \cite{Thiering2015}. The triplet state can be described as the occupation of $a_1$ and $b_1$ states  by parallel spins ($^3B_1$ state) whereas the optically allowed $^1B_1$ state was constructed from constrained occupation of symmetric orbitals [see Fig.~\ref{fig:OsExcitation}(d,e)]. By taking the excitation energy between the configuration of Fig.~\ref{fig:OsExcitation}(e) and (d), and the total energy difference between electron configuration of Fig.~\ref{fig:OsExcitation}(d) and the ground state singlet of Fig.~\ref{fig:OsExcitation}(b) we obtain the zero-phonon-line energy of 1.39~eV.  We assume an accuracy of about 0.2~eV of this procedure  based on our experience \cite{Thiering2015}. The shelving triplet state has about 0.04~eV higher energy than that of the ground state singlet. However, no low energy second triplet state exists near the optically allowed excited state singlet.
The spin-orbit coupling between $^1B_1$ and $^3B_1$ states is not allowed in the first order, thus the intersystem crossing is presumably very weak. On the other hand, if this scattering occurs then there should be an efficient and spin selective decay from the triplet shelving states to the ground state singlet because of the selection rule of $C_{2v}$ symmetry. Nevertheless, the optically detected magnetic resonance signal may be too weak to be detected because of the inefficient intersystem crossing between the optically excited state and the triplet shelving state. 

The negatively charged oxygen defects may only occur in phosphorus doped diamonds and thus we describe them only briefly. In $\mathrm{O_\text{S}}$($-$) the $a_1$ and $b_1$ orbitals are occupied by three electrons. HSE06 predicts an additional pseudo Jahn-Teller distortion from the $C_{2v}$ symmetry to $C_s$ symmetry (see Fig.~\ref{fig:sub}). However, the energy gain is only 0.01~eV. Motional averaging may result in an effective $C_{2v}$ symmetry in the measurements. In this effective symmetry the spin density is equally distributed in the two carbon dangling bonds whereas it results in 25\%-75\% distribution in the distorted structure. The $\mathrm{O_\text{S}}$($2-$) is a closed shell singlet in $C_{2v}$ symmetry where $a_1$ and $b_1$ states are fully occupied in both spin channels (see Fig.~\ref{fig:sub_KS}).

We note that our DFT calculations with an advanced functional reveal more complex nature of O$_\text{S}$ defect than that was reported by a previous local density functional approximation calculation carried out in a small supercell (Ref.~\onlinecite{Gali2001}), therefore, the conclusions of that previous study might be ambiguous.

\subsection{The oxygen-vacancy defect \label{sec:OV}}

The substitutional oxgen may combine with a vacancy forming the oxygen-vacancy (OV) defect. The OV defect may be taken as an analog with the famous NV center \cite{Davies1976} in diamond, except that oxygen brings one additional valence electron to the system. By following this argument one may imply that the electronic structure of OV($0$) may be similar to NV($-$) and it might be useful as a qubit but with enhanced T$_1$ time of the electron spin when the natural abundant $^{16}$O is involved in the defect with no nuclear spin; or rather $^{17}$O isotope is engineered in the OV defect with $I=5/2$ nuclear spin for enhanced number of quantum states for qubit operations. In this context it is important to note that one ESR center in CVD diamond samples was proposed to originate from OV($0$) defect but no spinpolarization could be detected in this ESR center upon photoexcitation \cite{hartland2014study}. Thus, we will particularly study the excitation of OV($0$) in a separate section (Sec.~\ref{sec:543.2}). In the rest of this section we analyze the electronic structure and the resulted magnetic properties of the OV defect.
 
\begin{figure}[ht]
\includegraphics[width=1.00\columnwidth]{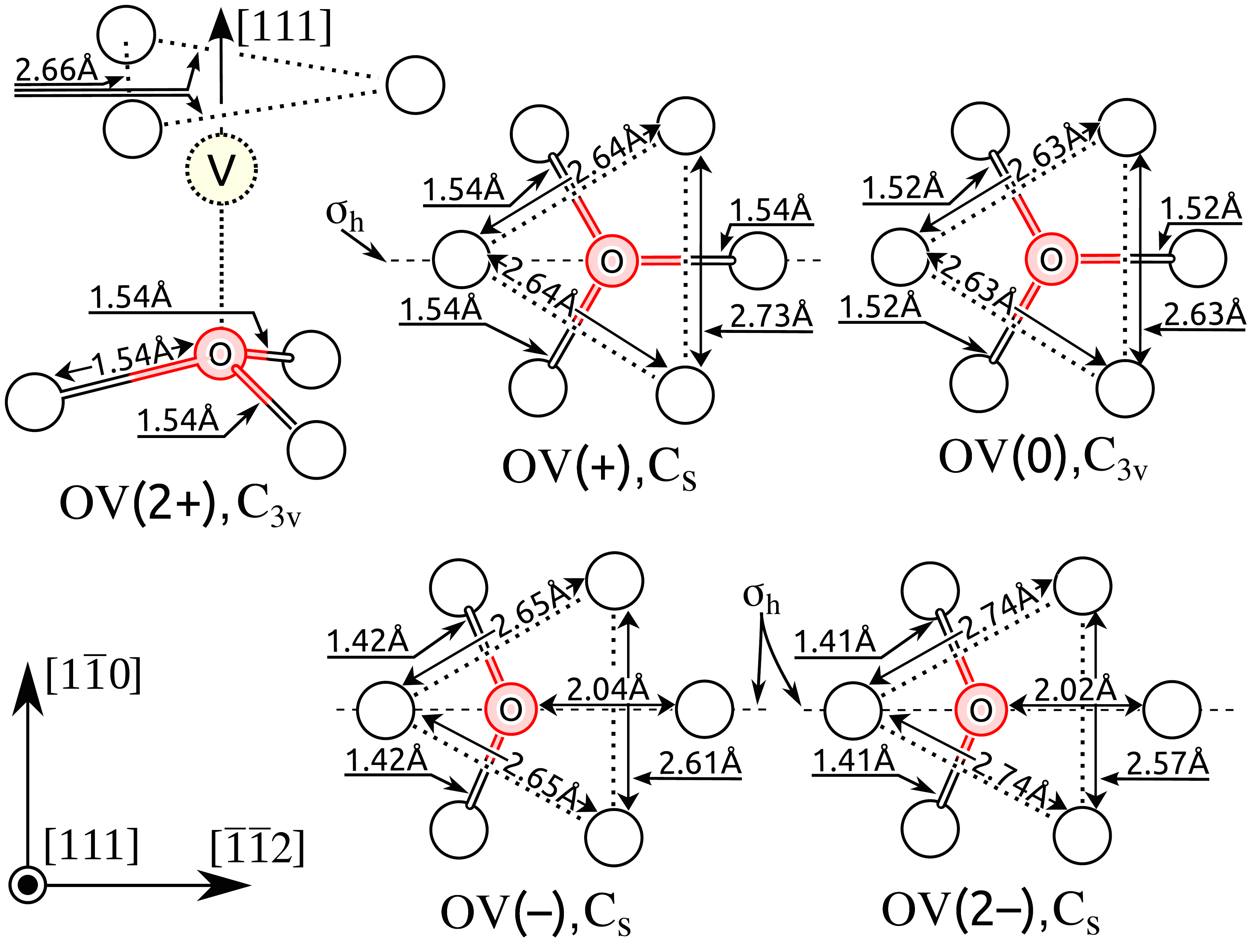}
\caption{(color online). 
Geometry of the $\mathrm{OV}$ defect in various charge states. The dotted circle labeled by V represents the position of the vacant site. $\sigma _h$ depicts the mirror plane.
}
\label{fig:vac}
\end{figure}

\begin{figure*}[ht]
\includegraphics[height=160px]{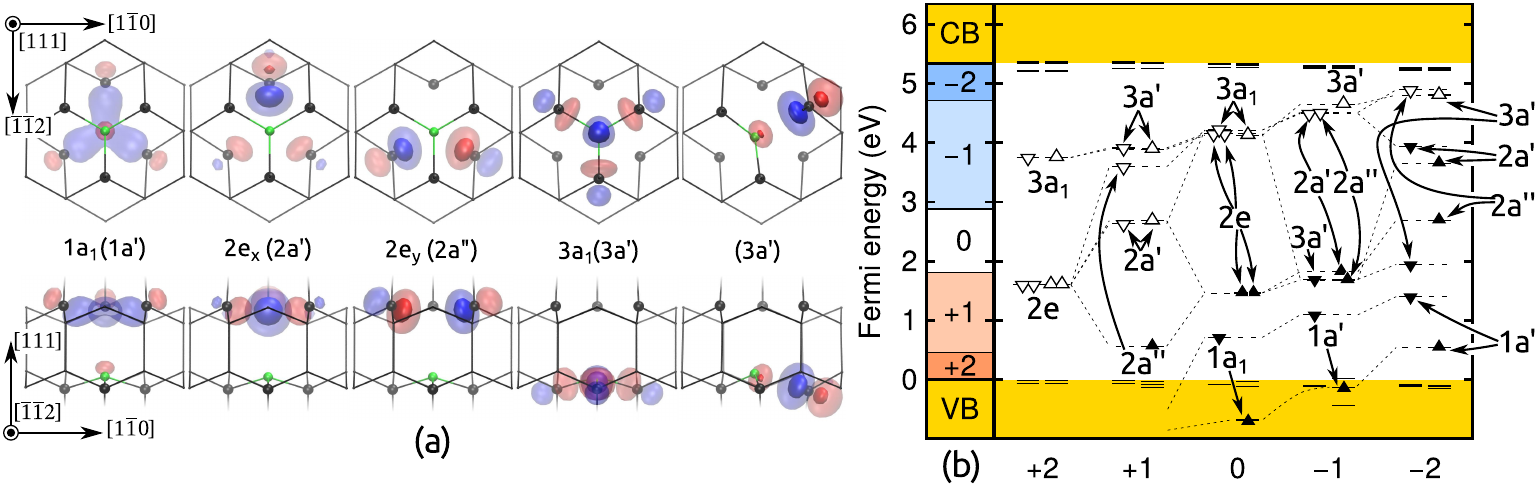}
\caption{(color online). 
Kohn-Sham (KS) orbitals (a) and levels (b) of the OV defect in various charged states. The charge transition levels are also given in (b). The KS orbitals are taken from the neutral charged state except for the last $(3a^\prime)$ orbital that is taken from the single negative charged state that leads to large distortion with breaking the $C_{3v}$ symmetry. The symmetry labels of orbitals in parentheses refer to the descendant orbitals in $C_s$ symmetry. 
The contour plots of the pseudo-wave functions are at the isovalues of $\pm 0.006$ (opaque) and $\pm 0.003$ (transparent).
}
\label{fig:OV_KS}
\end{figure*}

The undistorted OV defect has $C_{3v}$ symmetry (see Fig.~\ref{fig:vac}). In contrast to NV center in diamond possessing an $a_1$ and a degenerate $e$ level in the gap, OV defect exhibits three levels in the gap [see Fig.~\ref{fig:OV_KS}(b)]: an $1a_1$ level, a degenerate $e$ level and an additional $3a_1$ level. In the neutral charge state four electrons occupy these states, thus this system can be principally ionized. According to our results, the OV can form charge states from ($2+$) to ($2-$). The $1a_1$ level of the totally symmetric $1a_1$ orbital [see Fig.~\ref{fig:OV_KS}(a)] resides just above the valence band edge when the defect is neutral and it shifts to the valence band in the positive charged states. The $2e$ level comes from the degenerate $2e_x$ ($01\overline{1}$) and $2e_y$ ($2\overline{1}\overline{1}$) orbitals that are localized on the carbon dangling bonds near the vacancy. The third KS orbital is the antibonding $3a_1$ orbital of the three O-C bonds as shown in Fig.~\ref{fig:OV_KS}(a). If this state gets occupied in the negative charged states then the symmetry gets heavily distorted from $C_{3v}$. One of the three C-O bonds gets elongated (see Fig.~\ref{fig:vac}) where the extra electron will be localized. This resembles the fact that oxygen usually forms two covalent bonds. Next, we analyze the OV center in different charged states.

In the ($2+$) state, only the $1a_1$ state is fully occupied, and the remaining $2e$ and $3a_1$ state are empty. This defect exhibits $C_{3v}$ symmetry. This defect might only occur in B-doped diamond samples. The ($+$) state is a Jahn-Teller unstable system, thus its symmetry reduced to $C_s$, the $2a^{\prime\prime}$ is occupied in one spin channel from the former degenerate $e$ level. However, we note that this system is presumably a dynamic Jahn-Teller system as the calculated barrier energy of reorientation is tiny (10~meV) with showing an effective $C_{3v}$ symmetry. This system is analogous to the NV($0$): no ESR signal is expected to be observed in the ground state due to the dynamic Jahn-Teller effect.  

The neutral charge state is again a $C_{3v}$ defect with $S=1$ spin as the $e$ state is fully occupied in the spin majority channel which results in $^3A_2$ many-electron ground state. Thus, the OV($0$) system indeed looks similar to the NV($-$). The similarities for the ground state of these two defects are striking. The  calculated $D$-tensors due to electron spin - electron spin interaction are $D_{\text{NV}(-)}$=2848~MHz and $D_{\text{OV}(0)}$=2989~MHz. The calculated hyperfine tensors with proximate $^{13}$C nuclear spins of OV($0$) (see Table~\ref{tab:OV_hyper}) also follows closely those of NV($-$) defect in diamond \cite{szasz2013hyperfine}. Indeed, WAR5 ESR center in CVD diamond grown by oxygen chemistry yielded very similar spin-Hamiltonian parameters to those of NV($-$) defect \cite{cann2009magnetic_WAR5}: $D^\text{exp.}_{\text{NV}(-)}$=2872(2)~MHz \cite{NV_D1,NV_D2} and $D^\text{exp.}_{\text{WAR5}}$=2888(2)~MHz; the hyperfine parameters of WAR5 ESR center are shown in Table~\ref{tab:OV_hyper} and can be found in Refs.~\onlinecite{Gali2008, Felton2009, szasz2013hyperfine} for NV center. Based on these similarities it was proposed that WAR5 ESR center comes from OV($0$), although, there was no direct evidence for the presence of oxygen in WAR5 center \cite{hartland2014study} since the diamond was grown in the presence of natural abundance of oxygen isotopes. We find that the calculated hyperfine tensors of OV($0$) agrees well with the experimental data (see Table~\ref{tab:OV_hyper}). The interpretation of the experimental hyperfine satellites labeled as 3c/3h may be questioned. We believe that the hyperfine signals on 3c $^{13}$C isotopes may overlap with those of 3l $^{13}$C isotopes that makes the interpretation of these signals ambiguous. Nevertheless, the overall good agreement supports the identification of WAR5 ESR center as OV($0$). Definite experimental verification is possible in $^{17}$O enriched diamond samples where the hyperfine coupling with $^{17}$O $I=5/2$ spin should be well observed (around 15~MHz). 

\begin{table}[htbp]
\caption{Hyperfine tensors for the WAR5 ESR center (experimental data from Ref.~\onlinecite{cann2009magnetic_WAR5}) and the OV($0$) defect calculated by HSE06 functional in diamond. The direction of the hyperfine constants are given in angles in the parenthesis ($\vartheta$, $\varphi$). $\vartheta$ is the angle from [001], and $\varphi$ is the angle from [100] of the projection into the (001) plane. If no angle is given then the hyperfine interaction is axially symmetric and the high symmetry axis points towards [111]. The direction of $A$ constants for $^{13}$C isotopes are depicted in Fig.~\ref{fig:OV_hyper} with their corresponding labels (first column) where the $zz$, $yy$ and $xx$ components are shown by blue, green and red arrows, respectively. The number of symmetrically equivalent $^{13}$C isotopes are indicated in the first column preceding their label, and depicted in Fig.~\ref{fig:hyper_generic}.}
\begin{ruledtabular}
\begin{tabular}{clll}
  & $A_{zz}$ (MHz) & $A_{yy}$ (MHz) & $A_{xx}$ (MHz)  \\ 
\hline
WAR5 & $S=1$ & & \\
3a & 197.4(2) (54,225) & 117.3(2) (36,45)& 118.2(3) (90,135)\\ 
6g & 17.5(1) (60,225) & 11.7(1) (30,45)& 13.0(1) (90,135)   \\ 
3l &     12.6(1) (55,225) & 8.5(2)  (35,45)& 8.5(2) (90,135)    \\ 
3c/3h &    7.4(1) (55,225) & 4.3(1)  (35,45)& 4.3(1) (90,135)   \\ 
OV($0$) & $S=1$ & & \\
$\mathrm{^{17}O}$   & 16.5  & 17.9  & 17.9  \\ 
3a  & 188 (54,225) & 102 (36,45) & 102 (90,135)\\    
6g  & 17.6 (53,222) & 11.7 (133,176)& 11.8 (65,112)   \\ 
3l &    16.3 (57,225) & 10.9 (33,45)& 10.7 (90,135)    \\ 
3c &   -8.2 (85,45) & -9.8 (5,225)& -10.0 (90,135)   \\ 
6d  &   -3.4 (49,142) & -6.6  (127,93)& -7.0 (63,26)   \\ 
other  &   $<5$ & $<5$& $<5$  \\ 
3h  &   $<3$ & $<3$& $<3$  \\ 
\hline
OV($-$) & $S=3/2$ & & \\
$\mathrm{^{17}O}$  & -72.0 (120,45)  & -52.2 (150,225) & -51.1 (90,315)   \\ 
2a  & 130 (56,314) &  75.1 (92,43) & 74.7 (34,131)  \\     
1a  & 136 (126,45) & 74.5 (90,315) &  74.4 (144,225) \\   
1y  & 116 (55,225) & 36.7 (35,45) & 36.4 (90,315) \\   
other  &   $<13$ & $<13$& $<13$  \\ 
\end{tabular}   
\end{ruledtabular}
\label{tab:OV_hyper}
\end{table}

\begin{figure}[ht]
\includegraphics[height=160px]{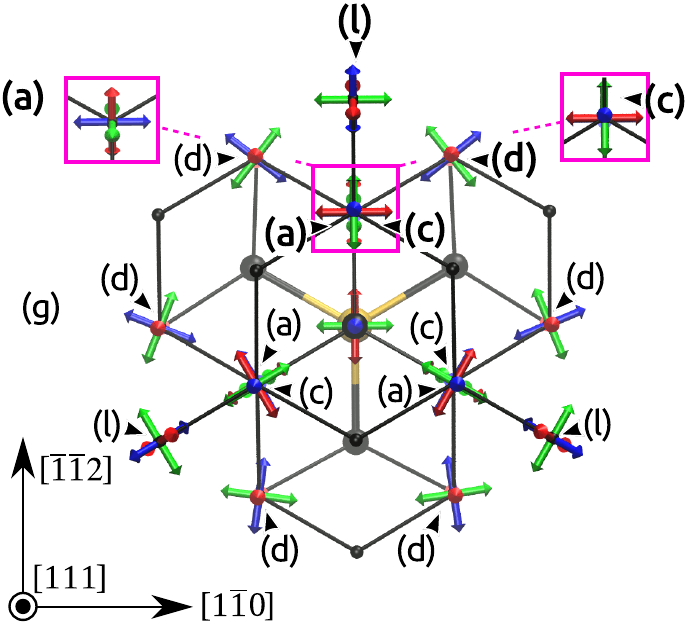}
\caption{(color online)
The direction of hyperfine eigenvectors of the OV($0$) defect where $zz$, $yy$ and $xx$ components are shown by blue, green and red arrows, respectively. The direction of hyperfine constants labeled by bold letters are in Table~\ref{tab:OV_hyper}. The yellow ball represents the O atom. For the sake of clarity, the (a) and (c) hyperfine tensors are shown separately in the two outlying purple boxes. The labeling of carbon atoms agrees with that in Figure \ref{fig:hyper_generic}.}
\label{fig:OV_hyper}
\end{figure}

We turn now to the negative charged state, OV($-$). The electron donated to the OV defect is localized in one of the C-O bonds and lowers the $C_{3v}$ symmetry to $C_s$ symmetry. As a consequence, the $e$ level will split and some of the corresponding states can strongly interact with each other resulting in highly correlated electron system which is very challenging for DFT modeling. We apply group theory analysis on the electronic states to estimate the total energy of the different eigenstates of the system. We concentrate on the open shell orbitals. Three electrons can occupy the $2a^{\prime}2a^{\prime\prime}3a^{\prime}$ orbitals (see Fig.~\ref{fig:OV_KS}) that results in $2^3=8$ possible configurations in total, including the spin degrees of freedom. After the reduction of the tensor product of three doublet orbitals,
\begin{equation}
\underset{(2e_{x})}{\underbrace{^{2}a^{\prime}}}\otimes\underset{(2e_{y})}{\underbrace{^{2}a^{\prime\prime}}}\otimes\underset{(3a_{1})}{\underbrace{^{2}a^{\prime}}}={}^{4}A^{\prime\prime}\oplus{}^{2}A^{\prime\prime}\oplus{}^{2}A^{\prime\prime}
\label{eq:OVm1_manybodywave}
\end{equation}
where the irreducible representation of the ascendant $C_{3v}$ symmetry is also shown in parenthesis. The resultant many-body states are two doublets and a quartet. Only the $|S=3/2, S_z=\pm 3/2\rangle$ states of the $^4A^{\prime\prime}$ can be described by
single Slater determinants; all the other states are multi-determinant in nature. In our Kohn-Sham DFT approach the following $S_z=1/2$ single determinant wave functions can be calculated: $\Phi_1=2a^{\prime}_\uparrow2a^{\prime\prime}_\uparrow3a^{\prime}
_\downarrow$,  $\Phi_2=2a^{\prime}_\uparrow2a^{\prime\prime}_\downarrow3a^{\prime}
_\uparrow$, $\Phi_3=2a^{\prime}_\downarrow2a^{\prime\prime}_\uparrow3a^{\prime}
_\uparrow$. The $\Phi_1$ assumes triplet coupling between the $2a^{\prime}$ and $2a^{\prime\prime}$ electrons, whereas the $\Phi_{2,3}$ suggests a singlet coupling. While the calculated DFT total energy of $\Phi_1$ is higher by 14~meV than that of $S=3/2$ state, the calculated difference between the total energies of $\Phi_2$ and $\Phi_3$ are even less than that, i.e.,  3~meV. It is important to notice that $\Phi_{1,2,3}$ states are not the true $S$ eigenstate of the system. However, the linear combination of  $\Phi_{1,2,3}$ states are true eigenstates of the $S_z \pm 1/2$ substates of the quartet (Eq.~\ref{eq:OVa}) and the doublets 
(Eqs.~\ref{eq:OVb}, \ref{eq:OVc}):
\begin{subequations}
\begin{alignat}{1}
    &\left|^{4}A_{1/2}^{\prime\prime}\right\rangle  =\frac{1}{\sqrt{3}}\left(\Psi_1+\Psi_2+\Psi_3\right) \label{eq:OVa}\\
    &\left|_a^{2}A_{1/2}^{\prime\prime}\right\rangle =\frac{1}{\sqrt{2}}\left(\Psi_{2}-\Psi_{3}\right) \label{eq:OVb}\\
    &\left|_b^{2}A_{1/2}^{\prime\prime}\right\rangle =\frac{1}{\sqrt{6}}\left(2\Psi_{1}-\Psi_{2}-\Psi_{3}\right) \label{eq:OVc}
\end{alignat}
\label{eq:OVeq}
\end{subequations}
The negligible total energy difference between $\Phi_{2,3}$ suggests that the interaction between the $3a^\prime$ and ($2a^\prime$, $2a^{\prime\prime}$) does not depend on flipping the spin between $2a^\prime$ and $2a^{\prime\prime}$. By neglecting the electron spin - electron spin and spin-orbit interaction energies and assuming that DFT total energies of $\Phi_{1,2,3}$ states are good expectation values one can estimate the total energies of the true eigenstates of the system by using the equations
\begin{subequations}
\begin{alignat}{1}
    &\left\langle \Psi_{1}\right|\hat{H}\left|\Psi_{1}\right\rangle =\Delta=\frac{1}{3}E\left(^{4}A^{\prime}\right)+\frac{2}{3}E\left(_b^{2}A^{\prime\prime}\right)\label{eq:OV2a}\\
    &\left\langle \Psi_{2,3}\right|\hat{H}\left|\Psi_{2,3}\right\rangle =\Lambda=\frac{1}{3}E\left(^{4}A^{\prime}\right)+\frac{1}{2}E\left(_a^{2}A^{\prime}\right)+\frac{1}{6}E\left(_b^{2}A^{\prime\prime}\right) 
    \label{eq:OV2b}
\end{alignat}
\label{eq:OV2}
\end{subequations}
By using HSE06 total energies we obtain $\Lambda$=14~meV and $\Delta=$288~meV. After solving Eq.~\ref{eq:OV2}, the total energy of $_a^{2}A^{\prime\prime}$ will be above that of the quartet by 0.57~eV.  The $_b^{2}A^{\prime\prime}$ state is still about 21~meV higher in energy than that of the quartet. Taking all the approximations into account this difference is within our accuracy, thus the ground state of OV($-$) can be either $^{4}A^{\prime\prime}$ or  $^{2}A^{\prime\prime}$. Both states are ESR active. The calculated hyperfine signals for the $^{4}A^{\prime\prime}$ are shown in Table~\ref{tab:OV_hyper}. The calculation of hyperfine tensors of the $^{2}A^{\prime\prime}$ configuration is not straightforward because of its correlated electronic structure.

The OV($2-$) defect is only marginally stable that might be found in P-doped diamond, thus we only briefly analyze it. We find that OV($2-$) also has multideterminant states, i.e., the $2a^{\prime\prime}_\downarrow$ and $3a^{\prime}_\uparrow$ holes in Fig.~\ref{fig:OV_KS}(b) form an open-shell singlet. The triplet state was favorable over the open-shell singlet by 7~meV which is within the accuracy of our method.

\subsection{Complexes of hydrogen with substitutional oxygen and oxygen-vacancy defect \label{sec:OH}}

Trapping of mobile interstitial hydrogen \cite{Goss_hydrogen_2002} by various point defects in diamond was already reported, such as the passivation of NV centers by hydrogen forming NVH complexes \cite{Stacey2012_Npassivation}, or vacancy-hydrogen complexes \cite{Glover2004, Shaw2005}. Silicon-hydrogen complexes were also found in diamond \cite{Goss_WAR3_2010, Thiering2015}. Here, we consider complex formation of hydrogen with substitutional oxygen and oxygen-vacancy defects. In substitutional oxygen defect some bonds are elongated that may be attacked by an approaching interstitial hydrogen. In oxygen-vacancy defect the carbon dangling bonds may be naturally saturated by hydrogen atoms. We study these complexes by our DFT machinery.

\begin{figure}[ht]
\includegraphics[height=160px]{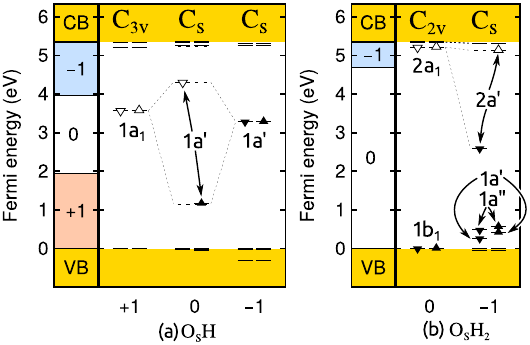}
\caption{(color online).
Kohn-Sham levels of the $\mathrm{O_\text{S}H_{1,2}}$ defects in their various charge states. The $1a_1$ ($\mathrm{O_\text{S}H}$) and $2a_1$ ($\mathrm{O_\text{S}H_{2}}$) state is localized in the symmetric combination of C-O antibonding orbitals with no saturation by hydrogen. The $1b_1$ state is localized in the two C-H-O bonds equally as well as the $1a^{\prime}$ and $1a^{\prime\prime}$ state in $\mathrm{O_\text{S}H_{2}}$ defect.
}
\label{fig:OsHn}
\end{figure}

First, we consider the saturation of the elongated C-O bond by a single hydrogen in O$_\text{S}$ defect with forming the O$_\text{S}$H defect. We find a single defect level occupied with a single electron [see Fig.~\ref{fig:OsHn}(a)], thus the defect can be positively and negatively charged. In the positive charge state this $a_1$ defect level is empty resulting in a high $C_{3v}$ symmetry. The $a_1$ state is the antibonding orbital of the C-O bonds. When this antibonding orbital is filled by an electron then it distorts the defect toward $C_s$ symmetry with $S=1/2$ paramagnetic state. We note that the calculated barrier energy for reorientation of hydrogen among the C-O bonds is 1.04~eV, thus no motional averaging is likely. The calculated hyperfine constants of $^{1}$H, $A_{xx}$=-0.9~MHz, $A_{yy}=$-7.4~MHz, $A_{zz}=$24.0~MHz, are very different from either the ESR signal in oxygenated HPHT diamond ($A_{||}$=35~MHz, $A_{\perp}$=32~MHz) \cite{Komarovskikh2013} or the ESR signal in CVD samples grown by oxygen chemistry ($A_{||}$=$\mp$13.6~MHz, $A_{\perp}$=$\pm$9.0~MHz) \cite{hartland2014study}. We conclude that ESR signal of this defect has not yet been detected. For future comparison, we report the hyperfine constants of $^{17}$O that are $A_{xx}$=-188~MHz, $A_{yy}$=-189~MHz, and $A_{zz}$=-224~MHz. A characteristic large $^{13}$C hyperfine coupling of $A_{xx}$=85~MHz, $A_{yy}$=85~MHz, and $A_{zz}$=331~MHz may be also detected. In the negative charge state the defect has a closed shell singlet ground state.

If two hydrogen atoms bind to $\mathrm{O_\text{S}}$ then the resultant $\mathrm{O_\text{S}}$H$_2$ defect will have $C_{2v}$ symmetry. The fully occupied $1b_1$ level lies just above the valence band edge whereas the empty $2a_1$ level is almost resonant with the conduction band edge. This state will be partially occupied in the negatively charged state of the defect with $S=1/2$ ground state but it is only marginally stable and may occur in P-doped diamond  [see Fig.~\ref{fig:OsHn}(b)]. The geometry distorts to $C_s$ symmetry because the unpaired electron resides in the antibonding state of C-O bonds. 

\begin{figure*}[ht]
\includegraphics[height=160px]{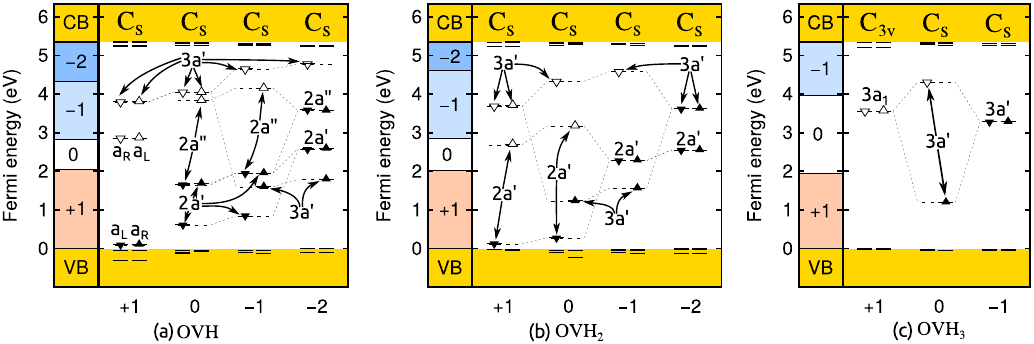}
\caption{(color online). 
Kohn-Sham (KS) levels of the complexes of hydrogen atoms and OV defect in various charge states. The KS orbitals are descendant from those of the OV defect (see Fig.~\ref{fig:OV_KS}), thus we follow the labeling the corresponding orbitals as those of OV defect. The labels $a_L$ and $a_R$ in (a) correspond to symmetry broken KS orbitals similar that of the O$_\text{S}$(0) system (see Fig.~\ref{fig:OsExcitation}) that leads to a correlated singlet ground state for OVH(+).}
\label{fig:OVHn}
\end{figure*}

Next, we consider the interaction between hydrogen and OV defect. Hydrogen much favors to bind to OV defect. OV defect has three carbon dangling bonds in the neutral charge state, thus we consider the complex formation of OV defect with one, two, and three hydrogen atoms. The calculated charge transition levels and the corresponding Kohn-Sham defect levels in the gap are depicted in Fig.~\ref{fig:OVHn}. All the defects are electrically active and can be paramagnetic at certain charged states that may be observed in ESR spectrum.

OVH defect may be positively charged in diamond when the Fermi-level is below $E_\text{V}$+2.1~eV.  The ground state of OVH($+$) is a correlated singlet
[see Fig.~\ref{fig:OVHn}(a)] that is more stable than the triplet coupling between by 0.16~eV within our hybrid DFT level. This effect is very similar to that of the $O_\text{S}$(0) that was shown in Sec.~\ref{sec:Os} and Fig.~\ref{fig:OsExcitation}. In this case two $^1A^\prime$ states correlate with each other (${2a^{\prime}}^2$) and (${2a^{\prime\prime}}^2$) but are unable to mix either with the first excited state $^1A^{\prime\prime}$ nor the triplet $^3A^{\prime\prime}$. The calculated ZPL of the optical transition between $^1A^\prime$ and $^1A^{\prime\prime}$ states is 1.38~eV where the relaxation energy upon optical excitation is 0.04~eV. This is a small value implying that the contribution of the phonons is not significant in the PL process.

We next focus our attention to the paramagnetic neutral OVH defect with $S=1/2$ ground state. The unpaired electron lies on the $2a^{\prime\prime}$ orbital [see Fig.~\ref{fig:OVHn}(a)]. This orbital originates from a degenerate $2e$ state of OV defect localized in the carbon dangling bonds (Fig.~\ref{fig:OV_KS}). This $2e$ state are filled by three electrons in OVH($0$) defect which is Jahn-Teller unstable in $C_{3v}$ symmetry and distorts to $C_s$ symmetry. As a consequence,  $2e$ level splits to the fully occupied $2a^{\prime}$ level and the half-occupied $2a^{\prime\prime}$ level. This situation resembles the case of NVH($-$) defect with $S=1/2$ spin state where the ESR signal of the $^1$H could be observed as motional averaged $C_{3v}$ symmetry.\cite{Glover2003}

\begin{figure}[ht]
\includegraphics[height=160px]{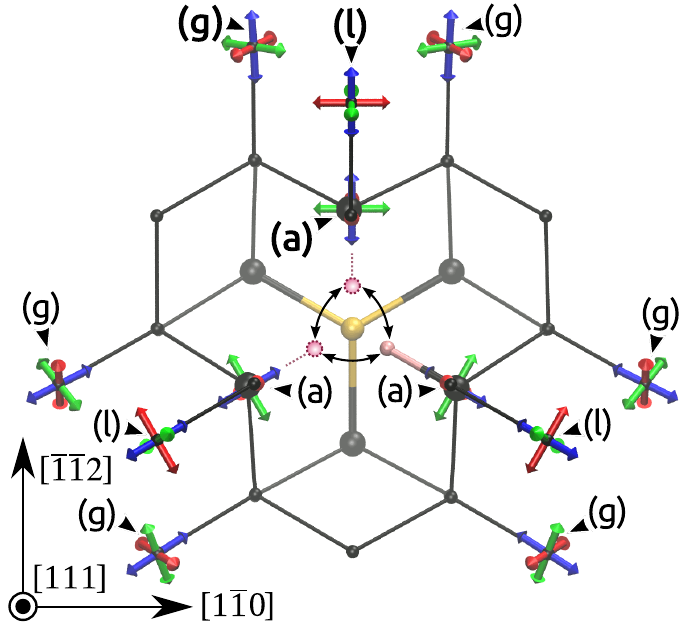}
\caption{(color online). 
The direction of hyperfine eigenvectors of the OVH($0$) defect where $zz$, $yy$ and $xx$ components are shown by blue, green and red arrows, respectively.  The actual angles of the bold labels are shown in Table~\ref{tab:OVH_hyper}. The pink ball represent the single motional averaged hydrogen atom. All hyperfine constants are obtained from the average of three equivalent configurations. The labeling of carbon atoms agrees with that in Figure \ref{fig:hyper_generic}.
}
\label{fig:OVH_hyper}
\end{figure}
We consider this effect for OVH($0$) defect as displayed in Fig.~\ref{fig:OVH_hyper}: the hydrogen may jump between the three equivalent $C_s$ configurations very rapidly, therefore ESR measurements will observe an effective $C_{3v}$ symmetry signal. This effect is discussed in detail in Sec.~\ref{sec:DJT} where we show by DFT calculations that the ESR signals of OVH(0) and NVH($-$) defects are motional averaged. Indeed, an ESR center was observed in CVD samples grown with oxygen chemistry that has $S=1/2$ spin state and one $^1$H hyperfine splitting with $C_{3v}$ symmetry that they associated with the OVH($0$) defect \cite{hartland2014study}, thus this center is labeled as OVH. The calculated motional averaged hyperfine signal of $^1$H is indeed very close to the experimental data of OVH ESR center, thus our calculation supports the identification of OVH($0$) defect in diamond. We provide the calculated hyperfine tensors of $^{17}$O  and proximate $^{13}$C nuclear spins that may be useful for future experiments. The ESR signal might be accompanied with an optical transition, thus we calculate the lowest possible excitation energy that corresponds to the transition of $2a^{\prime}\rightarrow2a^{\prime\prime}$. The calculated ZPL of the lowest energy optical transition is 1.11~eV where the relaxation energy upon optical excitation is 0.33~eV. The relatively small ZPL energy may imply that the non-radiative decay is very competitive and the luminescence is weak.

\begin{table}[htbp]
\caption{Hyperfine tensors for the OVH ESR center (experimental data from Ref.~\onlinecite{hartland2014study} labeled as exp.) and our theoretically predicted values for OVH$_n$ defects. We show the calculated static (labeled as static) and motional averaged (labeled as motional) values too. The direction of the hyperfine constants are given in angles in the parenthesis ($\vartheta$, $\varphi$). $\vartheta$ is the angle from [001], and $\varphi$ is the angle from [100] of the projection into the (001) plane. If no angle is given then the hyperfine interaction is axially symmetric and the high symmetry axis points towards [111]. The direction of $A$ constants for $^{13}$C isotopes are depicted in Fig.~\ref{fig:OVH_hyper} with their corresponding labels (first column) where the $zz$, $yy$ and $xx$ components are shown by blue, green and red arrows, respectively. The numbers on $^{13}$C isotopes refer to the symmetrically equivalent positions.}
\begin{ruledtabular}
\begin{tabular}{crlll}
  &  & $A_{zz}$ (MHz) & $A_{yy}$ (MHz) & $A_{xx}$ (MHz)  \\ 
\hline
\multicolumn{2}{c}{ OVH($0$)}  & $S=1/2$ & & \\
exp. & $\mathrm{^1H}$ & $\mp$13.6(1) & $\pm$9.0(1)  & $\pm$9.0(1)  \\ 
static    & $\mathrm{^1H}$          &  -17.1(130,238) & 13.7(45,270) & 1.6(73,163)  \\  
 & $\mathrm{^{17}O}$      & 23.0(116,209) & 24.0(45,270) & 25.2(56,138)  \\   
motional   & $\mathrm{^1H}$      & -16.3  & 7.2  & 7.2   \\   
  & $\mathrm{^{17}O}$  & 23.2 & 24.5  & 24.5   \\    
 3 (a) & $\mathrm{ ^{13}C}$  &   89(65,28) & 89(135,90)  & 175(55,136)  \\    
 6 (g) & $\mathrm{ ^{13}C}$  &   12(92,47) &  12(148,141) & 18(58,135)  \\     
 3 (l)  & $\mathrm{ ^{13}C}$ &   11(135,90)&  11(68,24)   & 17(53,132)  \\   
 other  & $\mathrm{ ^{13}C}$ &  $<5$ &  $<5$   & $<5$  \\
 \hline
 \multicolumn{2}{c}{ OVH($-$)}  & $S=1$ & & \\
  static    & $\mathrm{^1H}$          &  6.7(137,114) & -5.8(55,72) & -0.6(68,178)  \\ 
     & $\mathrm{^{17}O}$           & -115(128,139) & -86(72,64) & -85(136.8,-6.1)    \\  
  
 \hline
 \multicolumn{2}{c}{ OVH$_2$($+$)}   & $S=1/2$  &  &  \\ 
 static    & $\mathrm{^1H}$          &  26.3(85,135) & -25.8(96,45) & -23.5(7,11)  \\   
  & $\mathrm{^{17}O}$      & -3.7(45,88) & -1.0(135,90) & -1.0(91,-1)  \\   
 motional   & $\mathrm{^1H}$    & -24.6 & 0.9 & 0.9     \\  
   & $\mathrm{^{17}O}$  & -2.9 & -1.4  & -1.4  \\   
  3 (y) & $\mathrm{ ^{13}C}$  &   122(55,136) &  40(135,90)  & 40(65,27)  \\   
  other  & $\mathrm{ ^{13}C}$ &  $<10$ &  $<10$   & $<10$    \\
  \hline
   \multicolumn{2}{c}{ OVH$_2$($0$)}   & $S=1$  &  &  \\ 
static         & $\mathrm{^1H}$          &  -5.3(31, 20) & -11.1 (62.1,227) & 12.0 (78,131)  \\    
      & $\mathrm{^{17}O}$      & -76(135,90) & -79(70,21) & -107(52,128)  \\    
      
    \hline
  \multicolumn{2}{c}{ OVH$_2$($-$)}   & $S=1/2$  &  &  \\ 
      & $\mathrm{^1H}$          &  7.0(32, 69) & -2.9 (122,72) & -2.0 (88,161)  \\     
   & $\mathrm{^{17}O}$      & -260(54,135) & -206(135,90) & -205(66,26)  \\    
   1 (y) & $\mathrm{ ^{13}C}$  &   104(135,90) &  104(65,27)  & 343(55,136)  \\    
   other  & $\mathrm{ ^{13}C}$ &  $<15$ &  $<15$   & $<15$  \\
    \hline
    \multicolumn{2}{c}{ OVH$_3$($0$)}   & $S=1/2$  &  &  \\ 
    2 atoms    & $\mathrm{^1H}$          &  9.1(103,242) & -4.7 (105,148) & -3.2 (20,190)  \\  
    1 atom    & $\mathrm{^1H}$          &  5.0(86,45) & -2.6 (90,135) & -1.6 (176,45)  \\   
     & $\mathrm{^{17}O}$      & -229(126,45) & -180(36,45) & -180(90,135)  \\   
     1 (y) & $\mathrm{ ^{13}C}$  &   341(126,45) &  98(35,45)  & 98(90,135)  \\   
     other  & $\mathrm{ ^{13}C}$ &  $<30$ &  $<30$   & $<30$  \\
\end{tabular}       
\end{ruledtabular}
\label{tab:OVH_hyper}
\end{table}

The negatively charged OVH defect may be present in N-doped diamond samples. The extra electron occupies the $3a^{\prime}$ C-O antibonding orbital, thus the O atom creates two covalent bonds. Both the spin state and the symmetry of the neutral defect is ambiguous within the accuracy of our approach. The defect may form either $C_s$ or $C_1$ symmetry depending on the relative position of the oxygen and hydrogen impurities in the OVH defect. For the sake of simplicity we show the $C_s$ configuration in Fig.~\ref{fig:OVHn}(b). Two electrons may occupy the  $2a^{\prime\prime}$ and $3a^{\prime}$ orbitals either with antiparallel or parallel spins forming $S=0$ or $S=1$ states, respectively. We find that the $C_1$ $S=1$ configuration has the lowest total energy in our hybrid DFT approach but the total energy of the singlet states and the $C_s$ $S=1$ triplet are all within 30~meV that is beyond our accuracy. The hyperfine constants of the most stable triplet state is shown in Table~\ref{tab:OVH_hyper}. We note that motional averaging may occur in this defect that should alter the hyperfine constants of $^1$H and, with much lesser extent, that of $^{17}$O, as the reorientation of the geometry occurs between the carbon dangling bonds and the H atom of the defect, whereas the position of the oxygen atom remains almost intact.

The electronic structure of OVH$_2$ defect is similar to that of OVH. Out of three carbon dangling bonds two of them are terminated by H atoms in the OVH$_2$ system, thus only two defect states appear in the gap (see Fig.~\ref{fig:OVHn}). The first defect state is localized on the remaining carbon dangling bond ($2a^{\prime}$). The second level $3a^{\prime}$ is the antibonding orbital of the oxygen with its three neighbor carbon atoms. When this antibonding orbital gets occupied then one of three C-O bonds breaks and the initial geometry is heavily distorted. This occurs in the  ($0$), ($-$), and ($2-$) charge states. We predict that only the ($+$) and ($-$) charge states are ESR active with $S=1/2$ spin. We show that the calculated hyperfine tensors of the positive charge state in Table~\ref{tab:OVH_hyper}. Here we assume a motional averaging effect similar to that of the OVH($0$) with showing an effective $C_{3v}$ symmetry, but now there are six equivalent configurations instead of three because the H atoms can be interchanged. The neutral charge state is again either an open-shell singlet or a triplet. The singlet is preferred over the triplet by 19~meV, thus we cannot decide unambiguously the nature of the ground state. We suspect that the singlet state is indeed more favorable than the triplet one. The ($-$) state is a doublet, albeit its motional averaged hyperfine signal might be complicated. The H atoms may rotate among the 3 carbon dangling bonds but the rotation of the broken O-C bond may be hindered. As a consequence, the resultant hyperfine signal may possess only $C_s$ symmetry. We show the hyperfine parameters in $C_s$ symmetry in Table~\ref{tab:OVH_hyper}.

In OVH$_3$ defect the three dangling bonds are saturated, thus only the antibonding C-O $3a_1$ state appears in the gap. This defect level is occupied by one electron in the neutral charge state with $S=1/2$ spin. The corresponding hyperfine parameters are presented in Table~\ref{tab:OVH_hyper}. 

\section{Electron spin resonance spectrum of Jahn-Teller distorted systems  \label{sec:DJT}}

Defects that exhibit Jahn-Teller distortion may or may not show a high symmetry signal via motional averaging. This depends on the relative timescales of the microwave field applied to flip the electron spin in the ESR measurements and the tunneling between the symmetrically equivalent statically distorted structures. We first investigate this effect on OVH($0$) defect. In $C_{3v}$ symmetry the H sits in the symmetry axis of which total energy is 0.98~eV higher than that of the $C_{s}$ distorted geometry. This $\Delta _E$ energy is called Jahn-Teller energy. With using the nudge elastic band method \cite{Mills1995, Jonsson385nudgedelastic} we calculated also the barrier energy for reorientation among the symmetrically equivalent $C_{s}$ configurations that resulted in 0.93~eV. This is very close to the calculated Jahn-Teller energy. This Jahn-Teller system may be described as $E \times e$ type where $e$ now means $e$ vibration modes that dynamically drive the system out from $C_{3v}$ symmetry. In this system the tunneling rate $ \Gamma_E$ can be calculated  (see section 5.3 in Ref. \onlinecite{bersuker2013jahn} and section 4.3.3 in Ref. \onlinecite{bersuker2012vibronic}) as
\begin{subequations}
\begin{alignat}{1}
    & \Gamma_E = \frac{9 \kappa \Delta_E}{h}  \exp{\left(-\frac{6 \Delta_E}{\hbar \omega_A} \frac{\lambda}{1+3\lambda}\right)} \label{eq:eqTunnela} \\ 
    & \kappa = \left(\frac{16\lambda}{3\lambda^2+10\lambda+3}\right) ^2  \frac {9+54\lambda-6\lambda^3-\lambda^4}{2(9-\lambda^2)(1+3\lambda)^2}    \label{eq:eqTunnelb}\\
    & \lambda =\frac{\omega_B}{\omega_A} \label{eq:eqTunnelc} \text{,}
\end{alignat}
\label{eq:eqTunnel}
\end{subequations}
where $\lambda$ is defined by the vibration frequencies of the two modes $\omega_A$ and $\omega_B$ (not degenerate anymore) participating in the Jahn-Teller distortion as obtained in the $C_{s}$ configuration. In this particular system, $\omega_B$=0.18~eV is the bending mode of the C-H bond that reorients the system between two equivalent $C_{s}$ geometries whereas $\omega_A$=0.37~eV is the stretching mode of the C-H bond that drives the system through the $C_{3v}$ high symmetry configuration. In other words, $\omega_A$ and $\omega_B$ can be considered as the rotational and radial modes of the potential energy surface of the Jahn-Teller distorted defect, respectively.  We emphasize here that the values of $\omega_A$ and $\omega_B$ are comparable to that of $\Delta_E$. As a consequence, the calculated $ \Gamma_E$=25.5~THz tunneling rate is about three orders of magnitude faster than the X band at $\approx$10~GHz or the Q band at $\approx$34~GHz employed in ESR absorption measurements. We calculated the $\Delta_E$ and the reorientation barrier energy for the NVH($-$) defect too, and the calculated values at 0.96~eV and 0.90~eV are very similar to those obtained for OVH($0$) defect. The final result is $\Gamma_E$=26.6~THz for NVH($-$). This also demonstrates the similarities of these two defects and justifies our analysis on OVH($0$) defect as NVH($-$) defect was already proven to show motional averaged ESR signal.\cite{Glover2003} We note that our model takes the full complexity of the potential energy surface into account in the derivation of the tunneling rate, in contrast to a previous approach applied to NVH($-$) defect which only considered the radial modes.\cite{Kerridge2004} 

Based on these results we expect that all the OVH$_{1,2}$ defects exhibit $C_{3v}$ ESR signals because they also possess high C-H vibration frequencies that are comparable with the Jahn-Teller energy. 

We apply a similar analysis on O$_\text{S}$($+$) defect too. In this case the high symmetry $T_d$ is distorted to $C_{3v}$ symmetry that can be mediated by $t_2$ quasi-local vibration modes. The tunneling rate in this $T\times t$ system may be calculated \cite{bersuker2013jahn} as
\begin{equation}
    \Gamma_T =  \frac{2 \Delta_T}{h}  \exp{\left(-1.24 \frac{\Delta_T}{\hbar \omega_t } \right)} \text{,}
\label{eq:eqTunnel2}
\end{equation}
where $\Delta_T$ is the Jahn-Teller energy and $\omega_t$ is the corresponding vibration frequency. The calculated values are 0.76~eV and 0.067~eV, respectively, whereas the reorientation barrier energy is 0.51~eV. Despite the calculated Jahn-Teller and reorientation barrier energies are lower for O$_\text{S}$($+$) than those for OVH($0$) defect, the calculated $\Gamma_T$=0.3~GHz for O$_\text{S}$($+$) is much slower than $\Gamma_E$ is for OVH($0$). The reason of this difference is that the energy of the O-C related $t_2$ quasi-local vibration mode is much lower than $\Delta_T$. As a consequence, this rate is much slower than the frequency of the Q band (34~GHz) is that was employed in the ESR absorption measurements of KUL12 ESR center, thus the static $C_{3v}$ symmetry was detected. 

By comparing the cases of OVH($0$) and O$_\text{S}$($+$) it can concluded that by taking only the Jahn-Teller energies or the reorientation barrier energies can be misleading in the discussion of static versus motional average ESR signals of Jahn-Teller distorted defects but the ratio between the Jahn-Teller energy and the corresponding vibration energies that lead to the Jahn-Teller distortion should be considered in the analysis.

\section{Discussion of a luminescence center associated with oxygen and its  relation to the oxygen-vacancy defect \label{sec:543.2}}

In the CVD samples grown with oxygen chemistry a PL center was observed that has a characteristic zero-phonon-line (ZPL) at 543.2~nm \cite{hartland2014study} that corresponds to 2.28~eV. This 543.2-nm PL center and the WAR5 ESR center were detected in the same diamond sample, thus the 543.2-nm PL center was associated with OV($0$) defect \cite{hartland2014study}. As OV($0$) is isovalent with NV($-$), the optical spinpolarization of WAR5 ESR center was probed by photoexcitation at a wavelength that could excite the 543.2-nm PL center. However, no spinpolarization of OV($0$) was observed under illumination even at cryogenic conditions. We study the excited state of OV($0$) in details by means of \emph{ab initio} calculations, in order to reveal this issue.

\begin{figure}[ht]
\includegraphics[width=1.00\columnwidth]{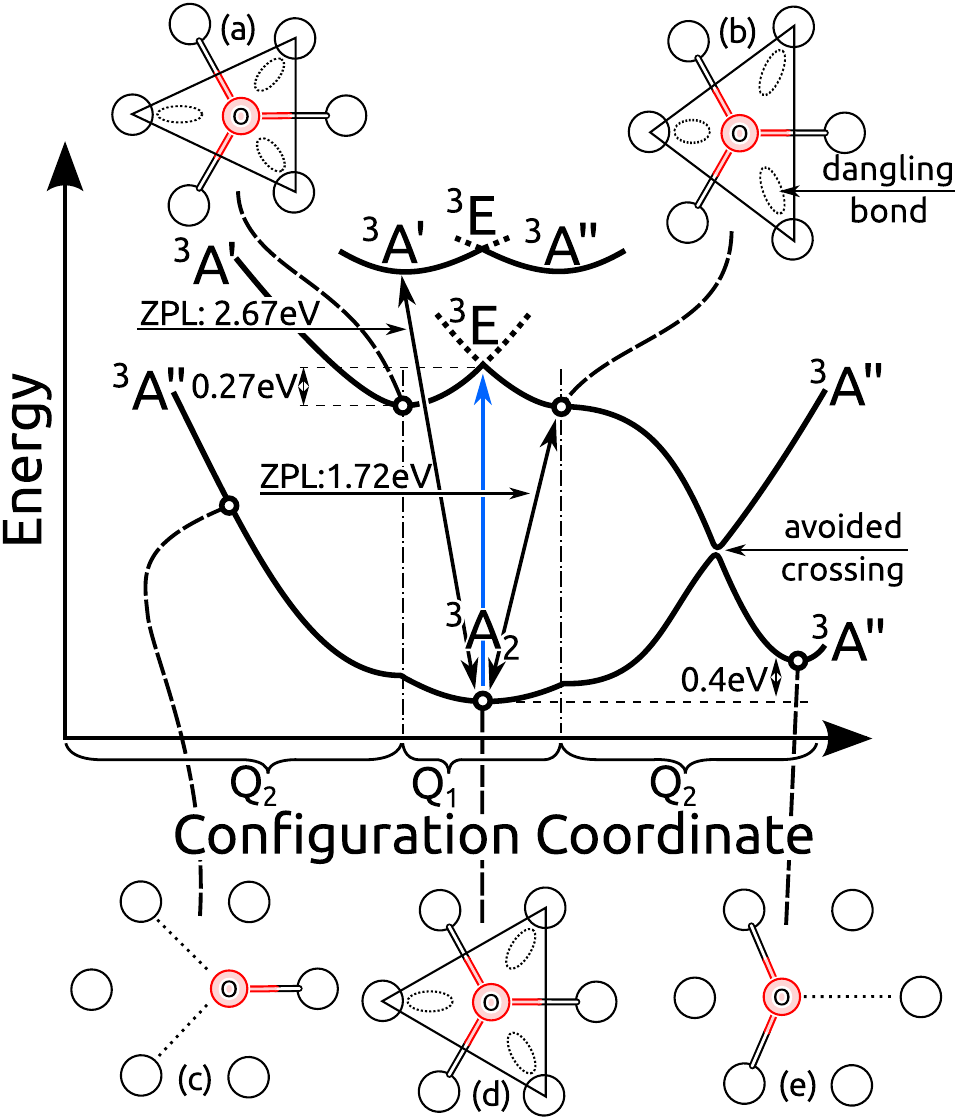}
\caption{(color online)
Potential energy surface of the ground and excited states of OV($0$). The different configurations of the defect are depicted. (d) shows the ground state with $C_{3v}$ symmetry. The oxygen atom forms covalent bonds with three carbon atoms and the carbon dangling bonds form an equilateral triangle. The vertical excitation leads to $^3E$ excited states (vertical arrow). In the lowest $^3E$ excited state the C-O antibonding orbital is occupied whereas only the dangling bonds are occupied in the higher $^3E$ excited state.  (a) $^3A^{\prime}$ type and (b) $^3A^{\prime\prime}$ type of distortion of the lowest $^3E$ excited state is depicted. The geometry is distorted along with a ($\mathrm{Q_1}$) normal coordinate, where the vacancy is distorted. The corresponding zero-phonon-line (ZPL) excitation energies belonging to the $^3E$ excited state configurations are given (athwart arrows). (c) Single C-O covalent bond and (e) two C-O covalent bond configurations are depicted. This type of distortion ($\mathrm{Q_2}$) can be characterized with the motion of the O atom. The latter forms a metastable state when the C-O antibonding orbital is occupied which leads to an avoided crossing with the same type of geometry in the ground state electronic configuration.}
\label{fig:OV_ex}
\end{figure}

The ground state of OV($0$) is indeed similar to NV($-$) where the $2e$ orbital is occupied by two electrons by parallel spins forming a $^3A_2$ ground state.  However, the excited state of the two defects may differ as the number of defect levels in the gap are different for the two defects. In the case of NV($-$) the only possible excitation is to promote an electron from the $1a_1$ level to the $2e$ level in the spin minority channel \cite{Gali2009}. In the case of OV($0$) one can promote an electron from the $2e$ level to the $3a_1$ level in the spin majority channel or from the $1a_1$ level to the $2e$ level in the spin minority channel. Our calculations predict that the former has lower ZPL excitation  energy (2.34~eV) than the latter (2.67~eV) by 0.33~eV. According to the Kasha-rule the electron from the higher excited state should decay to the lowest excited state rapidly by emitting phonons. Although, the symmetry of this excited state is $^3E$ but an antibonding C-O orbital is involved in it and not dangling bonds. Thus, the nature of this excited state significantly differs from that of NV($-$) defect. Obviously, the spin-orbit scattering rates from the excited state triplet to the singlet states are not analogous in the two defects. This $^3E$ state is Jahn-Teller unstable and the defect distorts to $C_s$ configurations as depicted in Figs.~\ref{fig:OV_ex}(a) and (b). The carbon dangling bonds of the vacancy may distort either to (a) $^3A^{\prime}$ state or (b) $^3A^{\prime\prime}$ state. The constraint DFT geometry optimization calculations predict $^3A^{\prime\prime}$ state to be more stable than $^3A^{\prime}$ state. 

We note here that the antibonding orbital is occupied in the ground state of OV($-$) that leads to the break of one C-O bond and a large reconstruction. Since this antibonding orbital is occupied in the lowest spin conserving excited state of OV($0$) we also study the configuration of strong reconstruction of OV($0$) in the excited state that leads to two C-O covalent bonds [see Fig.~\ref{fig:OV_ex}(e)] forming an  $^3A^{\prime\prime}$ state. This configuration is metastable with respect to the ground state of OV($0$) [Fig.~\ref{fig:OV_ex}(e)] as the total energy difference between the two states is only 0.4~eV. The configuration of one C-O covalent bond with breaking two C-O bonds [Fig.~\ref{fig:OV_ex}(c)] does not produce any metastable state.
 
\begin{figure}[ht]
\includegraphics[width=1.00\columnwidth]{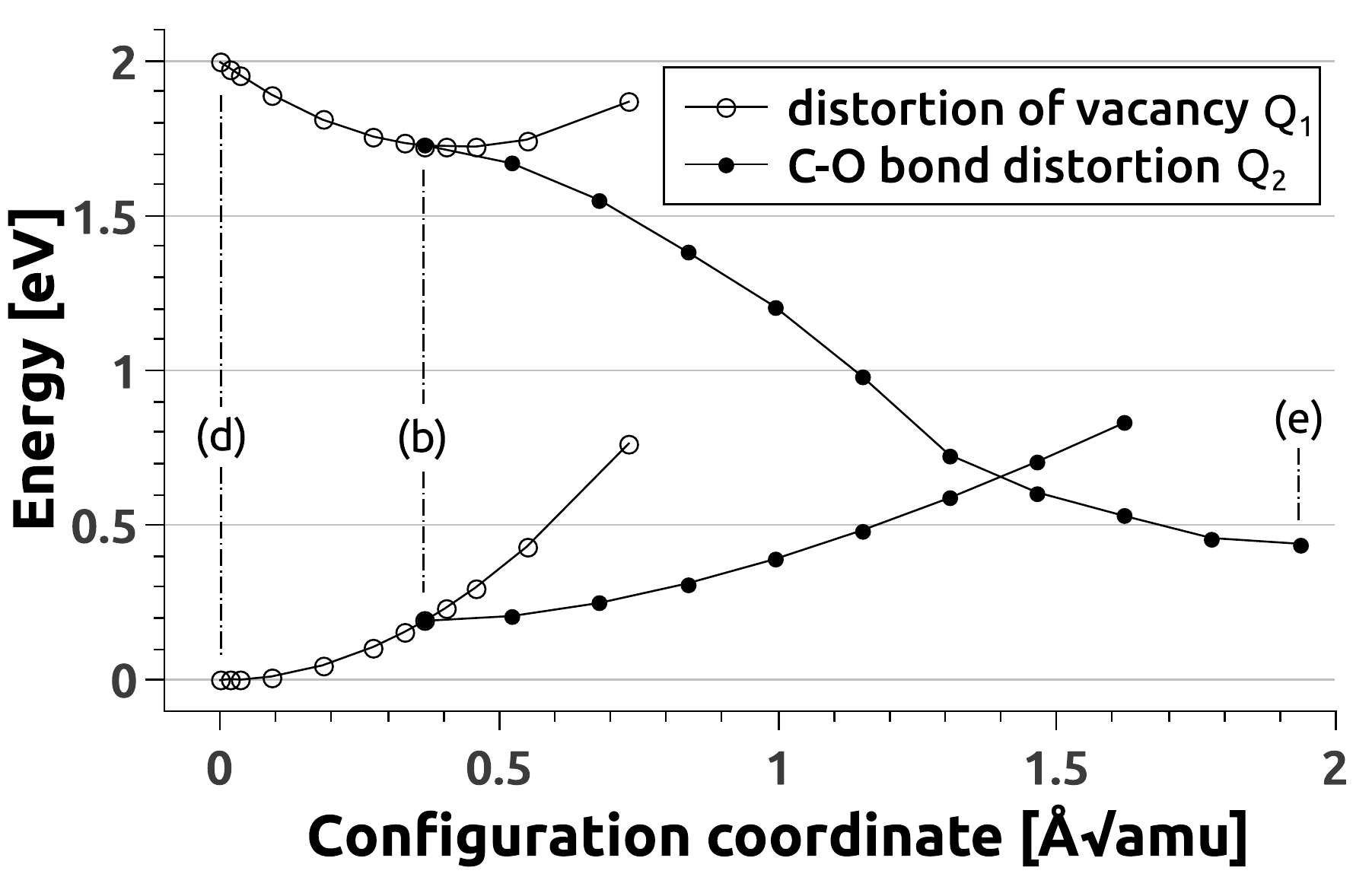}
\caption{(color online) The calculated potential energy surface of the ground and excited state of OV($0$) defect. The configuration coordinate ($\Delta d$) is measured with the formula $\Delta d=\sum_{i=1}^{3N} \sqrt{m_i} \Delta x_i$ where $N$ is the number of atoms, $m_i$ is mass of the corresponding atom in atomic mass unit ($m_u$) and $\Delta x_i$ is the motion of atoms with respect to the ground state geometry that is the chosen to be the reference. The reference in the energy is set to the total energy of the ground state. The empty circles ($\mathrm{Q_1}$ normal coordinate) represent the calculated data points when only the position of the carbon dangling bonds changes whereas the filled circles ($\mathrm{Q_2}$) show the calculated data points along the path leading to the break of one of the C-O bonds. The (d,b,e) letters denote the atomic configurations shown in Fig. \ref{fig:OV_ex}.
}
\label{fig:OV_pes}
\end{figure}

Our constrained DFT calculations do not lead a spontaneous decay from the lowest spin conserving excited state configuration Fig.~\ref{fig:OV_ex}(b) to the metastable configuration Fig.~\ref{fig:OV_ex}(e). Therefore we interpolate geometries between these two configurations and calculate the total energies. The potential energy surface plot is shown in Fig.~\ref{fig:OV_pes}. We find a virtually no-barrier decay from the excited state to the metastable state which has an avoided crossing with the ground state configuration. This result implies that the non-radiative decay is a very fast process and OV($0$) defect may have a very weak or no luminescence. We argue that the 543.2-nm PL center may not be associated with the OV($0$) defect despite the fact that the calculated ZPL energy (2.34~eV) is close to that of this PL center (2.28~eV). In any case, the spin-orbit couplings between triplet and the feasible multiplet singlet states do not follow those of NV($-$) defect because of the very different nature of the orbitals that build up the many-electron excited state. Our conclusion is that no optical spinpolarization of OV($0$) defect is expected in diamond which excludes OV($0$) from the family of NV-akin solid state qubits.			
		
The 543.2-nm PL center was observed in CVD diamond samples grown by oxygen chemistry where ESR centers associated with OV and OVH defects were observed that we support in our study. Thus, we consider the different charge states of OV defect and hydrogen complexes of OV defects as a candidate for 543.2-nm PL center. First, we calculate the excitation of OV($+$) defect by promoting an electron from the $1a^{\prime}$ level to the $2a^{\prime}$ level in the spin minority channel (see Fig.~\ref{fig:OV_KS}). In the excited state, the $1a^{\prime}$ hole orbital hybridizes with the valence band yet it has significant localized character. The calculated ZPL energy is 2.45~eV which is about 0.17~eV larger than that of the experimental value. We note that the calculated value is very approximative as the true excited state is multideterminant in nature that cannot be well described by our constraint DFT approach. Based on this argument OV($+$) is a feasible candidate for the 543.2-nm PL center.
								
Next, we consider the complexes of hydrogen and OV defect. The OVH($0$) exhibits an infrared optical transition. In the positive charge state, the lowest energy excitation can be described by promoting one electron from the valence band edge to the lowest energy empty $2a^\prime$ level in the gap. This energy is estimated by the calculated ($+|0$) charge transition level at 2.04~eV. This energy is smaller than the experimental value at 2.28~eV.  The calculated ($+|0$) charge transition level of OVH$_2$ at 2.02~eV may be considered but is again smaller than the experimental data. We note that this optical transition involves heavy geometry distortion and involvement of phonons in the PL process. The intra defect level transition energies of OVH$_{2,3}$ are too small to be a reasonable candidate. Finally, the calculated ($+|0$) charge transition level of OVH$_3$ (1.95~eV) is also too small.

In summary, we are not able to unambiguously associate any oxygen-related defects with the 543.2~nm PL center observed in CVD diamond grown with oxygen chemistry. We find that the OV($+$) is the best candidate among the considered defects. Further studies are needed to clarify this issue.

\section{Conclusion and summary \label{sec:Conclusion}}

We systematically characterized the interstitial and substitutional oxygen impurity and its complexes with a vacancy and hydrogen atoms. We identified the positively charged substitutional oxygen defect as KUL12 ESR center in oxygen implanted diamond. Our calculations support the relation of WAR5 and OVH ESR centers to neutral oxygen-vacancy and oxygen-vacancy-hydrogen complexes, respectively. Our results indicate the 543.2-nm PL center is not likely associated with the neutral oxygen-vacancy center. We rather predict that this defect has a very different excited state from that of NV center in diamond, and most probably exhibits a very fast non-radiative decay from the excited state. We predict that neutral OV defect will not act as NV-like qubit. We characterized the magneto-optical properties of paramagnetic oxygen defects in detail that may assist in finding oxygen defects in diamond for solid state quantum memory and metrology applications.

\acknowledgments
A.G. acknowledges the support from NIIF Supercomputer Center Grant No. 1090, EU FP7 Grant No.~611143 (DIADEMS), and the Lend\"ulet program of the Hungarian Academy of Sciences.


%

\end{document}